\documentclass[final]{siamltex}
\usepackage{times,algorithm,algorithmic,comment,epsfig,latexsym}

\def\AND{\wedge}
\def\OR{\vee}

\def\goesto{\rightarrow}
\def\implies{\Rightarrow}

\def\N{{\bf N}}
\def\cminus{\dot{-}} 
 
\def\PR{{\rm Pr}}
\def\HSAT{{\rm HORN}\hbox{-}{\rm SAT}}
\def\PUR{{\rm PUR}}
\def\beginproof{\noindent{\bf Proof.}\quad}
\def\endproof{}

\makeatletter
\newcommand{\singlespacing}{\let\CS=
        \@currsize\renewcommand{\baselinestretch}{1}\tiny\CS}
\newcommand{\singlespacingplus}{\let\CS=
        \@currsize\renewcommand{\baselinestretch}{1.15}\tiny\CS}
\newcommand{\doublespacing}{\let\CS=
        \@currsize\renewcommand{\baselinestretch}{1.75}\tiny\CS}
\newcommand{\draftspacing}{\let\CS=
        \@currsize\renewcommand{\baselinestretch}{2.0}\tiny\CS}

\makeatother

\def\desclabel#1{\bf #1\hfil}
\def\desc{\list{}{%
\labelwidth=\leftmargin
\advance \labelwidth by -\labelsep
\let \makelabel=\desclabel}}

\newtheorem{corrolary}{Corollary}
\newtheorem{example}{Example}
\newtheorem{observation}{Observation}
\newtheorem{claim}{Claim}

\def\qed{\hfill$\Box$\newline\vspace{5mm}}
\newenvironment{PROOF}{\noindent{\bf Proof:}}{{\qed}}

\author{Gabriel Istrate\thanks{
        Center for Nonlinear Science and CIC-3 Division,      
        Los Alamos National Laboratory,
        Los Alamos, NM 87545, 
        gistrate@cnls.lanl.gov}} 

\title{On the satisfiability of 
random $k$-Horn formulae}
\date{}
\begin{document}

\bibliographystyle{plain}

\maketitle
\begin{abstract} We determine the asymptotical satisfiability
probability of a random at-most-$k$-Horn formula, via a probabilistic 
analysis of a simple version, called \PUR, of positive unit resolution.
We show that for $k=k(n)\goesto \infty$ the problem
can be ``reduced'' to the case $k(n)=n$, that was solved in
\cite{istrate:cs.DS/9912001}. On 
the other hand, in the case $k=$ constant the behavior of \PUR\ is
modeled by a simple queuing chain, leading to a closed-form
solution when $k=2$. Our analysis predicts an ``easy-hard-easy''
pattern in this latter case. 
Under a rescaled parameter, the graphs of satisfaction probability
corresponding to finite values of $k$ converge to the one for the
uniform case, a ``dimension-dependent behavior'' similar to the one found
experimentally in \cite{kirkpatrick:selman:scaling} for $k$-SAT. 
The phenomenon is qualitatively explained by a threshold property for 
the number of 
iterations of \PUR\ makes on random {\em satisfiable} Horn
formulas. Also, for $k=2$ \PUR\ has a peak in its average complexity at
the critical point.    
\end{abstract}

\begin{keywords}
random Horn satisfiability, critical behavior, probabilistic analysis.
\end{keywords}
 
\begin{AMS}
68Q25,82B27
\end{AMS}
 
\pagestyle{myheadings}
\thispagestyle{plain}
\markboth{G. ISTRATE}{DIMENSION DEPENDENT BEHAVIOR OF RANDOM HORN 
SATISFIABILITY}

\section{Introduction}

Finding the ground state (state of minimum energy) of a physical
system and computing an optimal solution to a combinatorial 
optimization
problem 
are intuitively two very similar tasks. This simple observation, that
motivated the development of  {\em simulated annealing} 
\cite{simmulated:annealing}, a simple general-purpose heuristic for combinatorial
optimization, lies behind the
recent birth of a new field at the crossroads of Statistical
Mechanics, Theoretical Computer Science and Artificial Intelligence,
that studies {\em phase transitions in combinatorial problems} (see
\cite{hayes:cant:get:sat} for a readable introduction). The transfer of 
principles and
methods from Physics (mainly from Spin Glass Theory 
\cite{virasoro-parisi-mezard}) to
Computer Science has already been quite successful, and is responsible
for a couple of interesting results, such as a better understanding of
the factors that account for computational intractability
\cite{2+p:rsa, 2+p:nature},
strikingly accurate predictions of the average running time of various
algorithms \cite{scaling:search:cost:2,scaling:search:cost}, or of 
expected values of optimal solutions
\cite{mezard:parisi:matching}. 

The need for a rigorous validation of these insights is quite
obvious. The theory of spin glasses is a relatively young field, which
still presents many heuristic, unsolved or plain controversial aspects 
(for example see
\cite{non:mean:field:1,non:mean:field:2,non:mean:field:3} for a debate
on the validity and scope of the so-called Parisi solution of the
Sherrington--Kirkpatrick model). Moreover, while physical intuition can
guide the development of the theory for ``physical'' models, by corroborating (or 
falsifying) 
some of its predictions (e.g. see \cite{virasoro-parisi-mezard},
for a discussion of the demise, on physical grounds,
of the first formulation of the so-called {\em replica method}), such
intuition is not available when applying this type of ideas to 
combinatorial problems. Given that rigorous results are hard to come
by in the case of spin glasses proper, it is not surprising that while there has
been recently some progress (see e.g. 
\cite{talagrand:verres}), an analysis of most interesting 
combinatorial problems is still out of reach. 

An approach that was popular in Statistical Mechanics was to gather
intuition through the systematic study of {\em exactly solved models}
\cite{baxter:rigorous}. These are ``toy'' versions of the original models that
are simple to deal with, but retain much of the properties of the
former ones. We advocate such an approach for problems in 
Computer Science as well, and the purpose of this paper is to present
a (hopefully
nontrivial) ``exactly solvable satisfiability model'' that displays a 
{\em dimension-dependent behavior} fairly similar to the one observed 
previously in various
contexts such as percolation \cite{hara:slade:critical}, self-avoiding
walks, and recently for $k$-satisfiability by Kirkpatrick and Selman 
\cite{kirkpatrick:selman:scaling}. The problem
we investigate is {\em random Horn satisfiability}, and the 
``dimensionality'' of a formula is taken to be the 
{\em maximum length} of its clauses.\footnote{for technical 
convenience, all 
over the paper {\em random $k$-Horn satisfiability} is understood as 
{\em random {\bf at-most-$k$}-Horn satisfiability}.} 

\section{Overview}
There are actually two different notions of phase transition
in a combinatorial problem. The first of them, called 
{\em order-disorder phase transition} applies to optimization
problems and directly parallels the approach from Statistical 
Mechanics.
Potential solutions for an instance of $P$ are viewed as ``states'' of
a system. One defines an abstract {\em Hamiltonian (energy) function},
that measures the ``quality'' of a given solution, and applies methods
from the theory of spin glasses \cite{virasoro-parisi-mezard} to make 
predictions on the typical
structure of optimal solutions. In this setting a
phase transition is defined as non-analytical behavior of a certain
``order parameter'' called free energy,
and a discontinuity in this parameter, manifest by the sudden
emergence of a {\em backbone} of constrained ``degrees of freedom''
\cite{2+p:rsa} is responsible for the exponential slow-down of many 
natural algorithms.
 
The second definition is combinatorial and pertains to decision
problems. It relies on the concept of {\em threshold property} from
random graph theory, more precisely a restricted version of this
notion, called {\em sharp threshold}.
A satisfiability threshold always exists for monotone problems 
\cite{bollob-thomasson}, but may or may
not be sharp (we speak of a {\em coarse threshold} in the latter
case). 

The layout of the paper is as follows: in section~\ref{section:1} we 
review 
the results of Kirkpatrick and Selman, in particular discussing the 
concept of {\em critical behavior}, as well as some objectionable aspects 
of their results. 
  We then define the type of dimension 
dependent behavior we are interested in, argue that it captures to a 
large 
extent the results presented in \cite{kirkpatrick:selman:scaling}, and 
contrast it with  
critical behavior.  
Our results are presented and discussed in section~\ref{section:3}, 
while in
section~\ref{section:4} we further discuss their significance.

Finally for $k=2$, the one where the satisfaction probability has a 
singularity we are able to rigorously display another phenomenon that 
is 
believed to be characteristic of phase transitions: in many cases the 
``hardest on the average'' instances appear at the transition point 
(even if we only 
consider satisfiable instances \cite{achlioptas-sat-instances,mammen-hogg}); this feature is 
quite robust with respect to the choice of the particular algorithm 
\cite{cheeseman-kanefsky-taylor}. 
We are able to prove that for a {\em particular problem}, random 
at-most-2-Horn satisfiability,  the average 
running time of a {\em particular algorithm}, when restricted to 
satisfiable 
instances (the ones that are statistically significant on both sides of
the critical point) is finite outside the critical point, and it 
diverges as 
we approach this point, thus providing some evidence for the 
experimental 
wisdom.

\section{Phase transitions and critical behavior}\label{section:1}

We first discuss, briefly and limited to our interests, threshold
phenomena. Perhaps the best way to introduce them is through a concrete
example. To do this, we will use one ``canonical'' NP-complete
problem, {\em $k$-CNF satisfiability}. 

\begin{figure}
\label{sharp:thr}
\centerline{
\epsfig{figure=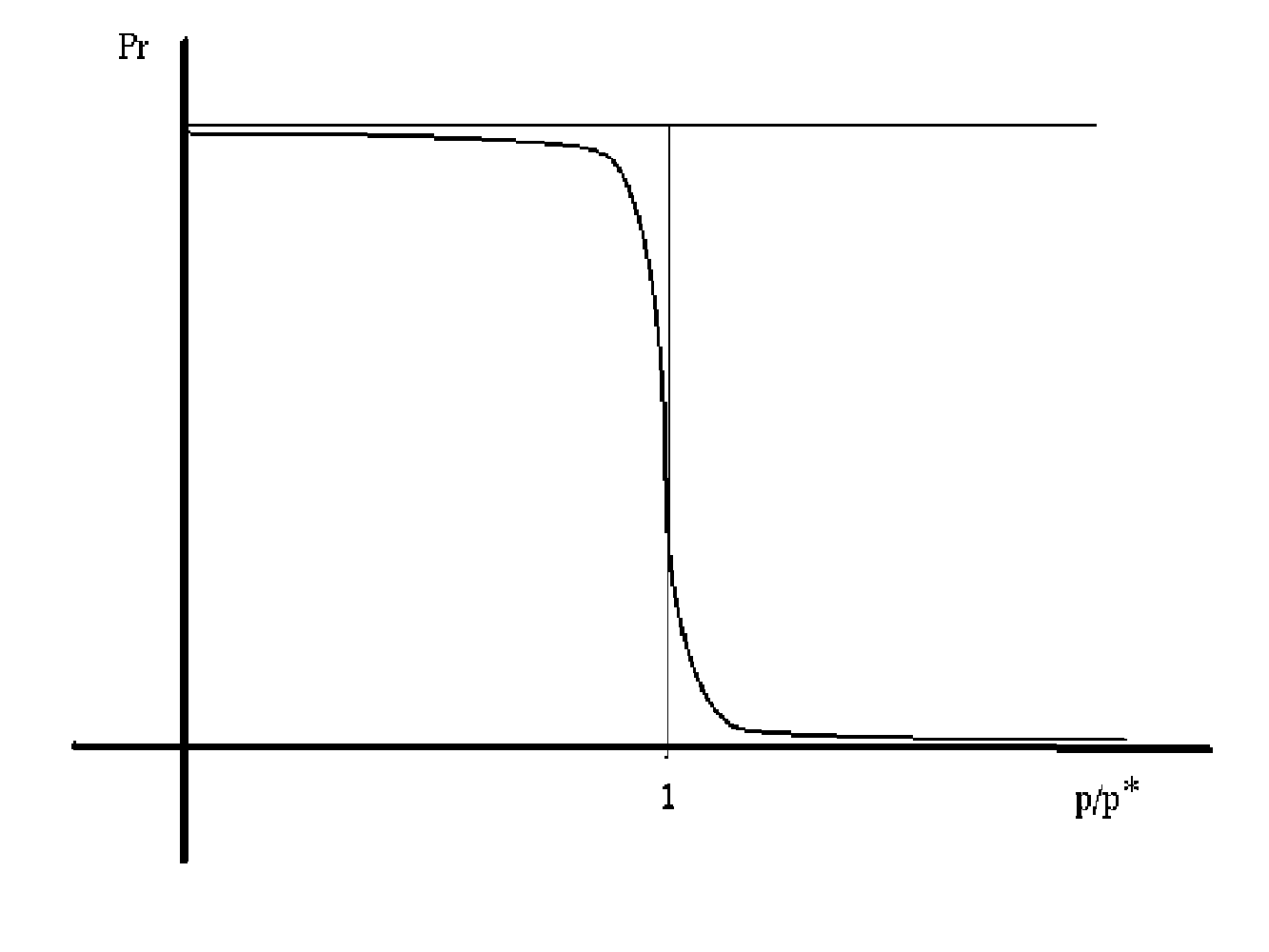,width=3.5in}}
\caption{Qualitative picture of a (rescaled) sharp threshold}
\end{figure}

To generate random formulas we use a 
model with one parameter, {\em the constraint
density $c$}, defined as the ratio between
the number of clauses $m$ and the number of variables $n$ of the
formula. A random formula is obtained by choosing $m$ random clauses. 
If we plot the probability that such a random formula is satisfiable 
against the constraint density $c$, we notice the existence of a
critical value $c_{k}$ such that the satisfaction probability drops
(as $n\goesto \infty$) from one to zero at $c_{k}$. Such a ``sudden
change'' is an illustration of the mathematical concept of {\em sharp
  threshold}, qualitatively illustrated in Figure~\ref{sharp:thr}. The
existence of a critical value $c_{k}$ has not been rigorously
established (except for $c_{2}=1$), even though Friedgut
\cite{friedgut:k:sat} has shown
that the transition is ``sharp'' for every $k$.  

Of special interest will also be the width of the so-called {\em scaling window (a.k.a. critical region)}. To define it consider, for $0 <\delta < 1$, 
$\alpha_-(n,\delta)$, the supremum over
$\alpha$ such that for $m=\alpha n$, the
probability of a random formula being
satisfiable is at least $1-\delta$.
Similarly, let
$\alpha_+(n,\delta)$ be the infimum over
$\alpha$ such that for $m=\alpha n$, the
probability of a random formula being
satisfiable is at most $\delta$.
Then, for $\alpha$  within the {\em $\delta$-scaling window}
\begin{equation}
W(n,\delta) = (\alpha_-(n,\delta),
\alpha_+(n,\delta)),
\end{equation}
the probability that
a random formula is satisfiable is
between $\delta$ and $1-\delta$.

We will be interested in the width of the window
$W(n,\delta)$ as a function of
$n$. It is generally believed that $|W(n)|=\theta(n^{-1/\nu})$
for some $\nu=\nu_{k}\geq 1$ independent of $\delta$, even though the existence of $\nu_{k}$ has
only been established for $k=2$ \cite{scaling:window:2sat}.  

\subsection{Order/disorder phase transitions}

Statistical mechanics deals with the description of systems having a 
large 
number of degrees of freedom. One of its fundamental predictions 
concerns the 
fact that at thermal equilibrium each such state occurs with 
probability 
proportional to $exp(-\beta H(\sigma))$, where $\beta$ is an {\em 
inverse 
temperature}, and $H$ is a {\em Hamiltonian function}, describing the 
energy of 
the particular state $\sigma$. The resulting distribution is called 
{\em 
the Gibbs distribution $G_{\beta}$} given by 
\[
\Pr[\sigma]=\frac{exp(-\beta\cdot H(\Phi;\sigma))}{Z[\Phi]},
\]
where
\[
Z[\Phi]= \sum_{\sigma\in \{0,1\}^{n}}exp(-\beta\cdot H(\Phi;\sigma))
\]
is the so-called {\em partition function}.
 
Changes in the order properties of the system, 
which characterize order-disorder phase transitions, manifest 
themselves as  
non-analytical behavior of thermal averages (i.e. averages over the 
Gibbs distribution) of a certain {\em order parameter}.
We want to emphasize that the physicists' use of the term order
parameter would be quite different from the one from combinatorics. 
An order parameter is a quantity that is zero on one side of the 
phase transition and becomes non-zero on the other side (for instance 
the satisfaction probability could be an order parameter).   
 
One of the simplest illustrations of these 
concepts is the {\em two-dimensional Ising model} (see 
\cite{baxter:rigorous} for a 
thorough treatment).  In this model we
have a number of {\em spins}, that are small magnets located on the
vertices of the two-dimensional lattice, and pointing either                   
up or down. The spins interact with their neighbors and with an {\em 
external
magnetic field $h\in {\bf R}$}, which will tend to align the spins in one of the
two directions. The energy of a state $\sigma$ is 
\[
H(\sigma)=- \sum_{i\sim j}\sigma_{i}\cdot \sigma_{j} + h\cdot 
\left(\sum_{i}\sigma_{i}\right).
\]

The order parameter is called {\em free energy}, is a function of
temperature, and is formally defined as 

\[
f = -\frac{1}{\beta n} \ln Z[\Phi].
\]

It measures the fraction of spins that are ``frozen'' 
when the field is turned off. 

We now briefly describe the essence of the phase transition: 
above a certain temperature $T_{c}$, {\em the Curie-Weiss point}, when 
the magnetic field is turned to zero
the proportion of spins that point in each direction is about
$\frac{1}{2}$ (the so-called {\em disordered phase}). But for
temperatures below $T_{c}$ when we turn the field to zero some
orientation still dominates (the  {\em ordered phase}), and the proportion of
spins pointing up(down) changes discontinuously as $h$ passes through zero.

 The connection with combinatorial optimization follows from the
observation that when $\beta \goesto \infty$ (that is the temperature
approaches 0 K), the Gibbs distribution $G_{\beta}$ converges to a
uniform distribution $G$ on the set of states of minimal energy
(ground states). Thus, based on  this analogy, one can hope that 
ideas from Statistical Mechanics are able to provide insight into the 
structure of optimal solutions to an instance of a problem in 
Combinatorial Optimization. Rather than providing a complete discussion (which 
would require to 
rigorously define the notion of optimization problem) we will discuss 
this in the
context of MAX 3-SAT, the optimization version of satisfiability. For 
now 
it suffices to mention the three main ingredients of an optimization 
problem, 
its {\em instances}, {\em solutions} to instances of a problem, and an 
{\em 
cost function}, that measures the quality of a solution for a certain 
instance. 

\begin{example}(MAX 3-SAT)

{\bf Input:} A propositional formula $\Phi$ in conjunctive normal form, 
such that every
clause has length exactly 3.
 
{\bf Solution:} A truth assignment $\sigma$ for the propositional 
variables in $\Phi$
that maximizes the number of satisfied clauses.
 
{\bf Cost function:} The cost $C(\Phi,\sigma)$ of a truth assignment 
$\sigma$ for an instance 
$\Phi$ of MAX 3-SAT is the number of clauses of $\Phi$ that are 
violated by 
$\sigma$. 
\end{example}
 
Let $Q$ be an optimization problem and let  $\Phi$ be an instance of 
$Q$ ``on
$n$ variables'' (i.e., all solutions have length $n$). We view the
set of all assignments on $\{0,1\}^{n}$ as ``states of a system.'' To 
each such
state $\sigma$ we associate the Hamiltonian (energy function)
\[
H(\Phi;\sigma)=\mbox{ the cost of instance }(\Phi;\sigma)\mbox{ of }Q.
\]                                                                              
\begin{example}
Let $\Phi$ be a 3-CNF formula, and let $\sigma$ be an assignment. 
According 
to the previous definition $H(\Phi;\sigma)=C(\Phi;\sigma)$. $H$ can be 
formally expressed \cite{monasson:zecchina} as
 
\[
H(\Phi;\sigma)=\sum_{l=1}^{m}\delta\left[\sum_{i=1}^{n}C_{l,i}\cdot 
(-1)^{\sigma_{i}};-3\right],
\]
\end{example}
where $\delta[i;j]= 1_{\{i=j\}}$ is the Kronecker symbol and $C_{l,i}$ 
is 1 if the $l$th
clause contains the literal $x_{i}$, $-1$ if it contains 
$\overline{x_{i}}$ and
zero otherwise.

For the case of problems of interest to Computer Science the instance           
$\Phi$ is not fixed, but rather is a sample from a certain
distribution. This is very similar to the context of {\em spin-glass 
theory}, 
a subfield of Statistical Mechanics. 
The extra ingredient of this theory is that the coupling coefficients 
are no 
longer considered fixed, but are rather independent samples from a 
certain 
distribution. In the language of the theory of spin glasses  $\Phi$ is 
called a {\em quenched quantity}).
 
As in the case of the Ising model, the order parameter 
is the {\em ground state free energy}, more precise its expected value 
\[
\overline{f}=-\frac{1}{\beta n}\overline{\ln(Z)},
\]
where $\overline{(\ldots)}$ stands for the average over the random
distribution of $\Phi$.                                                         
\begin{definition}
A {\em physical (order/disorder)
phase transition} in a combinatorial optimization problem
is a point where $\overline{f}$ is not analytical.
\end{definition}

Free energy has  an especially crisp intuitive
interpretation in the case of the problem MAX 3-SAT 
\cite{monasson:zecchina}:
 
\begin{example}\label{3sat:expl}
Let $\Phi_{n}$ be an instance of MAX 3-SAT, let $A$ be the set of 
optimal
assignments to $\Phi_{n}$, endowed with the uniform measure $\mu_{n}$.
Statistical Mechanics predicts that, as $n\goesto \infty$, $\mu_{n}$ is
``close'' to a product measure on $\{0,1\}^{n}$, $\mu_{1,n} \ldots
\mu_{n,n}$. The {\em free energy per site} $f$ is the fraction of
variables $x_{i}$ that are (asymptotically) {\em fully constrained} 
(that is
$\mu_{i,n}$ converges in distribution to a measure having all its
weight on one of the two points 0,1.                                           
\end{example}

\section{Critical behavior and the mean-field 
approximation}\label{section:2}

An important feature that order/disorder
phase transition share with the combinatorial notion of {\em threshold 
properties} (that are usually the type of phase transition of interest 
in combinatorics) is that the various quantities of interest,
such as the satisfaction probability, the ground state energy, and the
location of the phase transition are hard to compute. No
general-purpose  methods exist, and in some cases even obtaining good
non-rigorous estimates is a challenging open problem. 

A technique  that often provides realistic approximate values for
these quantities
came to be known as the {\em mean-field (annealed) approximation}. In a nutshell 
a mean-field approximation assumes that we are trying to compute the
average (over a certain discrete probability space) of a certain
expression $f\circ (g_{1}, \ldots, g_{n})$. Then the mean 
field-approximation amounts to taking 

\[
E[f(g_{1}(x),\ldots, g_{n}(x)]\sim f[E[g_{1}(x)],\ldots, E[g_{n}(x)]]. 
\]

This technical definition of the mean-field approximation does not
convey a useful intuition: suppose we want to solve a  
combinatorial problem whose objective function depends on 
simultaneously satisfy several ``constraints'' whose effects are 
usually not independent. The mean-field approximation ignores
 the dependencies between various constraints, and treat them
as independent.  
\vspace{5mm}
\begin{example}
Let us return to the case of spin glasses. Each configuration of spins 
$\sigma$
has an energy specified by a {\em Hamiltonian} $H(\sigma)$. A typical
expression for $H(\sigma)$ is 
\[
H(\sigma)=\sum_{i\sim j}a_{i,j}\sigma_{i}\sigma_{j},
\]

where the $a_{i,j}$'s are interaction coefficients between adjacent 
spins
(according to some adjacency graph specific to the considered model).
The quantity of interest, {\em average free energy} $\overline{f}$
is hard to compute directly because of the logarithmic function present 
in 
the definition of the free energy. In this context the mean-field
approximation amounts to 

\[
\overline{f}\sim -\frac{1}{\beta n}\ln[ \overline{Z[\Phi]}]. 
\]
\end{example}
\vspace{5mm}

The advantage of this heuristic is that the average on the right-hand 
side is one that is usually much easier
to compute.

For combinatorial phase transitions, the mean-field approach usually
amounts to an approximation using the so-called {\em first-moment
method} 

\vspace{5mm}

\begin{example} {\bf ($k$-Satisfiability)}

The reason that the satisfiability probability of a random formula is
hard to compute is that, for two assignments $A,B$ the events $A\models
\Phi$ and $B\models\Phi$ are not independent. 
One way to construct a mean-field theory for $k$-SAT is to ignore the
dependencies between these events.  More precisely, we have

\[ 1_{SAT}[\Phi] = f(g_{A_{1}}[\Phi], \ldots, g_{A_{2^{n}}}[\Phi]),
\]
where 
\[ f(x_{1}, x_{2}, \ldots, x_{2^{n}})= 1 - \prod _{i=1}^{2^{n}}x_{i},
\]
and 

\[ g_{A}[\Phi]= \left \{\begin{array}{ll}
                 1, & \mbox{ if }A\not \models \Phi,
                 \\
 
                 0,  & \mbox{ otherwise.}\\ 
        \end{array}
\right.
\]
Define $\gamma_{k}=1-2^{-k}$. The mean-field approximation amounts to 
\[ 
\Pr[\Phi \in SAT] = E[1_{SAT}[\Phi]]\sim f(E_{g_{1}}[\Phi], \ldots,
E_{g_{2^{n}}}[\Phi])
\]
Since 

\[
E_{g_{1}}[\Phi]= \ldots = E_{g_{2^{n}}}[\Phi])= 1-\gamma_{k}^{cn}
\]
this reads,
\[ 
\Pr[\Phi \in SAT]\sim 1- \left[1-\gamma_{k}^{cn}\right]^{2^{n}}\sim 
1-e^{-
2^{n}\cdot \gamma_{k}^{cn}}= 1- e^{-E[\#_{SAT}[\Phi]]}
\]
where $\#_{SAT}[\Phi]$ is the number of satisfying assignments for
$\Phi$. Thus (neglecting the case $E[\#_{SAT}[\Phi]]=1$) 

\[ 
\Pr[\Phi \in SAT]= \left \{\begin{array}{ll}
                 1, & \mbox{ if }E[\#_{SAT}[\Phi]]\goesto \infty,
                 \\
 
                 0,  & \mbox{ if }E[\#_{SAT}[\Phi]]\goesto 0.\\ 
\end{array}
\right. 
\]

\end{example}
\vspace{5mm}
\subsection{Critical exponents and behavior}

A phenomenon that has been observed in various contexts is 
{\em critical behavior}. In these cases the class of problems under
study has an intrinsic notion of 
dimensionality $d$, and in the limit $d\goesto \infty$ (or sometimes 
even when $d$ is greater than a so-called {\em critical dimension}) 
``the annealed approximation becomes exact''.

A way to give precise meaning to the above quote comes from the
concept of {\em universality}. In Statistical Mechanics one define
certain {\em critical exponents}, that describe the behavior of the
system near the critical points; universality predicts that phase 
transitions
with the same critical exponents are ``structurally similar''. 

Since critical exponents can be defined for the mean-field versions of
the physical models too, critical behavior means that as $d\goesto 
\infty$
(or, sometimes, for $d$ larger than a value called {\em the upper 
critical dimension}) the critical
exponents of the $d$-dimensional system coincide with the critical
exponents of the $d$-dimensional mean-field model.   

\vspace{5mm}
\begin{example}
{\bf (Bond) percolation on the lattice ${\bf Z}^{d}$.}
Percolation \cite{grimmett:percolation} is a mathematical theory that 
models the flow of  liquids in random porous media. In our case 
the flow is on the
lattice ${\bf Z}^{d}$ of dimension $d$, and the model has one
parameter, the edge probability $p\in [0,1]$. Each bond (grid
edge of the lattice ${\bf Z}^{d}$) is considered open with
probability $p$ (independently of the other bonds) and the order
parameter is the probability $P_{d}(p)$ that the origin lies in an
infinite cluster. $P_{d}$ is a monotonically increasing function of
$p$. It is  believed that $P_{d}(p)$ is  zero up to a {\em critical 
value $p_{c}(d)$} (known
rigorously only for $d=2$), greater than zero beyond that point, and
non-analytical but continuous (at least for $d=2$) at $p_{c}(d)$.
It is also believed that above 
(and around the critical value) $P_{d}(p)\sim (p-p_{c}(d))^{\beta}$ where $\beta$ is a 
{\em critical exponent} that depends on $d$ but {\em not} on
the explicit lattice considered (i.e. it would be the same if we choose
another $d$-dimensional lattice instead of ${\bf Z}^{d}$). This is 
only one of the several critical exponents that are believed to
structurally characterize percolation on $d$-dimensional lattices (see
\cite{grimmett:percolation}).  

Without going into further details, we note that 
the ``mean-field approximation''
corresponds to considering percolation on the {\em $d$-dimensional
  Bethe lattice}, a
nd the critical behavior 
amounts to the observation that for $d$ greater than a {\em critical
dimension} (known to be at most 16 \cite{hara:slade:critical}, and is 
believed to be 6) the 
critical exponents of percolation on ${\bf Z}^{d}$ are those of
percolation on the Bethe lattice.   

\end{example}

\subsection{Rescaling and critical behavior}
\label{discuss}
A recent example of critical behavior has recently been observed
experimentally by Kirkpatrick and Selman
\cite{kirkpatrick:selman:scaling} for satisfiability problems. 

Their results does not mention 
critical exponents (although it is closely related).  To explain 
them, 
we need to 
introduce first another concept from Statistical Mechanics: {\em 
finite-size
scaling}. The intuition behind it is that
\cite{kirkpatrick:selman:scaling} ``sufficiently close to a threshold
or critical point, systems of all sizes are indistinguishable except
for an overall change of scale.'' In
mathematical terms this amounts to defining a new order parameter
that ``opens up'' the {\em scaling window, 
the region where the probability decreases from 1 to 0.}  
\vspace{5mm}
\begin{example} {\bf Hamiltonian Cycle.}

The random model has one parameter $m$, the number of edges. A random
sample is obtained by choosing uniformly at random a set of $m$
distinct edges of a complete graph with $n$ vertices. The following 
result (obtained by Koml\'{o}s and Szemer\'{e}di \cite{hamcyclerand}) 
describes the phase
transition in this problem: 

Let $m=m(n)= \frac{1}{2}n\cdot \log(n)+\frac{1}{2}n\cdot \log 
\log(n)+c_{n}\cdot n$. Then

\[
\lim_{n\implies \infty}Pr[G\mbox{ has a Hamiltonian cycle}]=\left 
 \{\begin{array}{ll}
                  0, & \mbox{ if $c_{n}\goesto -\infty$,}
                 \\
                 e^{-e^{-2c}},  & \mbox{if $c_{n}\goesto c$,}\\
                 1, & \mbox{ if $c_{n}\goesto \infty$.}
                 \\  
        \end{array}
\right.
\] 

A rescaled parameter for the Hamiltonian cycle problem can be defined
by $c_{n}=\frac{1}{n}\cdot [m-\frac{1}{2}n\cdot
\log(n)-\frac{1}{2}n\cdot \log \log(n)]$. This parameter yields a
rescaled limit probability function $f(c)=e^{-e^{-2c}}$. 
\end{example}
\vspace{5mm}

It is important to note that, since an annealed approximation yields 
an expression for the order parameter (in our case satisfaction
probability) that will usually display a phase transition as well, 
a rescaled parameter can be defined for the mean-field version of the 
problem as well. 

The definition of the rescaled parameter allows a precise formulation
of the intuition that an annealed approximation becomes exact in the
limit $d\goesto \infty$. Let $P_{d}$ be a class of satisfiability
problems indexed by a dimensionality parameter $d$, let $F_{d}$
be the rescaled satisfaction probability graph of $P_{d}$, and let 
$F_{ann,d}$ be 
the rescaled graph corresponding to the annealed approximation. 
Kirkpatrick and Selman observe experimentally that {\em as $d\goesto 
\infty$, 
the function sequences $F_{d}$, $F_{ann,d}$ converge punctually to a 
common limit $F_{\infty}$}. 
\vspace{5mm}
\begin{example}
We present in detail the experimental results of Kirkpatrick 
and Selman. They define an (approximate) rescaled parameter for $k$-SAT
\[
y_{k} = n^{1/\nu_{k}}\frac{(c-c_{k})}{c_{k}},
\]
where $c=m/n$, $c_{k}$ is the critical threshold for $k$-SAT, and
$\nu_{k}$ is the scaling width coefficient.  
Also, define the ``annealed rescaled parameter'' 
\[
y_{\infty,k} = n\frac{(c-c_{k})}{c_{k}},
\]

The rescaled limit probability graphs (and, see below, the rescaled
versions of the mean-field versions) seem to converge (see Fig. 4 in
that paper) to the ``annealed limit'' 

\[
f_{\infty}(y) = e^{-2^{-y}}. 
\]
\end{example}
\vspace{5mm}

\vspace{5mm}
\begin{definition}
In this paper {\em dimension-dependent 
behavior} refers to the above-mentioned type phenomenon, convergence 
of the ``rescaled'' probability functions (and their annealed
counterparts) to some common {\em annealed limit}.
\end{definition}

\vspace{5mm}
\begin{observation}

It is important to note that dimension-dependent behavior is at the
same time more and less demanding than critical behavior.

It is more demanding since it requires that the 
annealed approximation be exact {\em throughout the (rescaled version) 
of
the critical region}. In contrast, critical exponents only provide a
qualitative picture of this region, rather than uniquely determine the
limit probability throughout it; for instance the width of the scaling 
window $\nu$ is equal to $2\beta+\gamma$, where  $\beta$ 
is the
so-called {\em order-parameter exponent}, that characterizes the
asymptotic behavior of the order parameter close to the transition
point, and $\gamma$ is called {\em susceptibility exponent} (see
e.g. \cite{scaling:window:2sat}). 

It is less demanding since it does
not assume the existence of critical exponents, therefore 
{\em it makes sense for problems having coarse thresholds,
including those that have no singular/critical points}.

\end{observation}
\vspace{5mm}

Why should we expect critical behavior and the above form for the 
annealed 
limit ? The intuition is very simple: the major difficulty in computing 
the 
probability that a random $k-SAT$ formula is satisfiable is the fact 
that, for two assignments $A$ and $B$, the events ``$A\models \Phi$'' 
and 
``$B\models \Phi$'' are not generally independent, because there exist 
clauses of length $k$ that are falsified by both $A$ and $B$. On the 
other hand, 
qualitatively, as $k\goesto \infty$ clausal constraints become 
progressively ``looser'', so that in the limit we can neglect such 
correlations.

As to the exact expression for $f_{\infty}(y)$, for a $k$-CNF formula 
the mean-field approximation implies

\[
\Pr[\Phi \in \overline{SAT}]\sim (1-\gamma_{k}^{cn})^{2^n}\sim 
e^{-2^{n}\cdot \gamma_{k}^{cn}}. 
\]

But since $c_{k}$ is specified (in the mean-field approximation) by 
$E[\# SAT]\sim 1$, i.e. $2^{n}\cdot \gamma_{k}^{c_{k}n}\sim 1$, 
or $1+c_{k}\log_{2}\gamma_{k}=0$, this implies that as $k\goesto 
\infty$ 
\[
\Pr[\Phi \in \overline{SAT}]\sim e^{-2^{n\cdot [1-c/c_{k}]}}\sim 
f_{\infty}(y_{\infty,k}).
\]

In other words, when plotted against the annealed order parameters 
$y_{ann,k}$ the rescaled satisfaction probability graphs (and their 
annealed 
counterparts) punctually converge to the graph of $f_{\infty}$. 

\section{Does critical behavior really exist ?}

The intuitive argument sketched in the preceding paragraph seems to provide 
a beautiful explanation of the experimental results from \cite{kirkpatrick:selman:scaling}. That this 
intuition is, however, problematic has been shown by Wilson 
\cite{wilson:ksat:wrong}. First 
note that if the previous argument were true, we would have 
$\nu_{k}=1$ 
for any large enough $k$, since this is the width of the scaling
window that the mean-field versions of $k-SAT$ predict.  
On the other hand Wilson 
presented a simple argument that implies that $\nu_{k}\geq 2$)
Hence the above explanation is not rigorously valid.  

We stress that Wilson's observation does {\em not} rule out the 
existence of critical behavior: we, in fact, believe that the 
qualitative intuition that motivated \cite{kirkpatrick:selman:scaling}, 
that versions of $k-SAT$ become more and more ``similar'' as $k$ goes to 
infinity, is correct. {\em It is the notion of annealed approximation that 
needs to be changed}.  
And, certainly, {\bf his results do not rule the possibility that the rescaled 
limit probabilities converge, as $k\goesto \infty$, to a 
suitable-defined limit}. Obtaining a rigorous example where this holds, 
that identifies a 
suitable ``annealed approximation that becomes exact'' and also obtains 
an 
explanation for this convergence,  could hopefully  
offer insights on how to address this problem 
for random $k-SAT$ as well. This is what our theorems in the next section 
provide.   

\section{Our results}\label{section:3}
A {\em Horn clause} is a disjunction of literals containing {\em at
most one positive literal}. It will be called {\em positive} if it
contains a positive literal and {\em negative} otherwise.
A Horn formula is a conjunction of Horn
clauses. {\em Horn satisfiability} (denoted by $\HSAT$) is the
problem of deciding whether a given Horn formula has a satisfying
assignment.                                                

In this chapter we prove a result that displays 
dimension-dependent behavior for (at most) $k$-Horn satisfiability, the 
natural version of Horn
satisfiability studied, parameterized by the maximum clause length.  
This problem is also of practical
interest in Artificial Intelligence, 
mainly in connection to {\em theory approximation}
\cite{kautz-selman-kc}. 
The results can be summarized as
 follows:
 
\begin{enumerate}
\item For an unbounded $k=k(n)$ the threshold phenomenon
is essentially the one from the ``uniform case'' $k(n)=n$. 
In particular there exists a
``rescaled'' parameter that makes the graphs of the limit probabilities 
superimpose (Theorem~\ref{k:infinite}). 
 
\item For any constant $k$ the threshold phenomenon is qualitatively
described by a suitably chosen queuing model
(Theorem~\ref{k:3etc}). This yields a 
closed-form expression for the satisfaction probability when 
$k=2$ (Theorem~\ref{k:2}). This expression has a singularity (though $k=2$
is likely the only case that does so).  
\item The rescaled limit probabilities from the
cases when $k$ is a constant converge to the one from the ``infinite'' 
case, that can in turn be seen as the result of a mean-field approximation
(thus the problem displays what we have called dimension-dependent behavior). 

\item Somewhat surprisingly, the explanation for this convergence (an
intrinsic feature of the problem) is
a threshold property for the number of iterations of PUR
(a particular algorithm) on random satisfiable Horn formulas 
``in the critical range.'' 

\item In the case when $k=2$ \PUR\ displays an 
``easy-hard-easy'' pattern for the average number of iterations on 
satisfiable instances, peaked at the point where the limit probability
has a singularity (Theorem~\ref{k:2:runtime}). 
\end{enumerate}
\vspace{5mm}

Note, however, the important difference between 
random $k$-SAT and random at-most-$k$-\HSAT: for every $k\geq 2$,
$k$-SAT has a sharp threshold
\cite{friedgut:k:sat}. All versions of \HSAT\ have coarse thresholds.

\vspace{5mm}
\begin{definition}
Let $k=k(n):\N \goesto \N$ be monotonically increasing, $1\leq
k(n)\leq n$. We define the following random model $\Omega(k,n,m)$:
{\em formula $\Phi$ on $n$ variables 
is obtained by selecting (uniformly at random
and with repetition) $m$ clauses from the set of all (non-empty) Horn
clauses in the given variables of length {\em at most $k(n)$}.}
\end{definition}
\vspace{5mm}

The following are our results (whose proofs are only sketched):
\vspace{5mm}

\begin{theorem} \label{k:infinite}
If $k(n)\goesto \infty$, $c>0$, 
$H_{k(n)}$ is the number of Horn clauses on $n$ variables
having length at most $k(n)$,  and $m(n)= c\cdot \frac{H_{k(n)}}{n}$ 
then 
\begin{equation}
\label{formula:1}
p_{\infty}(c):=\lim_{n\goesto \infty} Pr_{\Phi \in 
\Omega(k(n),n,m)}(\Phi \in \mbox{HORN-SAT}\/) =
1-F_{1}(e^{-c}).
\end{equation}
\end{theorem}

\vspace{5mm}

\begin{theorem}\label{k:2}
If $c>0$, and $F_{2}:(0,1)\goesto (1,\infty)$,
$F_{2}(x)=\ln x/(x-1)$, then 
\begin{equation}
\label{formula:2}
p_{2}(c):=\lim_{n\goesto \infty} Pr_{\Phi \in \Omega(2,n,cn)}(\Phi \in 
\mbox{HORN-SAT}\/) =
\left \{\begin{array}{ll}
                 1, & \mbox{ if $c\leq \frac{3}{2}$,}
                 \\
 
                 F_{2}^{-1}(2c/3),  & \mbox{ otherwise.}\\ 
        \end{array}
\right.
\end{equation} 
\end{theorem}
\vspace{5mm}

More generally, define $\lambda_{k}=\frac{k!}{k+1}$ and
$S_{j}^{i}={{i}\choose {0}}+{{i}\choose {1}}+\ldots+{{i}\choose
{j}}$ (with the usual convention ${{i}\choose{j}}=0$ for $i<j$). Then

\begin{theorem}\label{k:3etc}
The limit probability $p_{k}(c):=\lim_{n\goesto \infty}
Pr_{\Phi \in \Omega(k,n,c\cdot n^{k-1})}(\Phi \in \mbox{HORN-SAT}\/)$
is equal to the probability that the following Markov chain
ever hits state zero:
\begin{equation}\label{eq:3etc}
\left \{\begin{array}{l}
        Q_{0}=1,\\
        Q_{i+1}=Q_{i}\cminus 1+Po(c\cdot \lambda_{k}\cdot 
S_{k-2}^{i+1}),\\
\end{array}
\right.
\end{equation}
\end{theorem}
\vspace{5mm}

To get a better intuition on the threshold phenomenon, as displayed by
Theorems~\ref{k:infinite}, \ref{k:2} and \ref{k:3etc}, we have plotted
(in Fig. 1) the limit probability functions 
$p_{2}(\cdot),p_{3}(\cdot),p_{\infty}(\cdot)$, against the ``rescaled'' parameter 
(inspired by Theorem~\ref{k:infinite}) $\hat{c}=\frac{m\cdot
n}{H_{k(n)}}$. This rescaling has the pleasant property that it
simplifies the factor $\lambda_{k}$ from the right-hand side
of~\ref{eq:3etc}, in particular mapping the critical point in
Theorem~\ref{k:2} to $\hat{c}=1$. 
The graphs of $p_{2}$ (continuous) and $p_{\infty}$
(dashed) are obtained from their formulas in the previous results,
while $p_{3}$ (dotted) is obtained via simulations. The figure makes
apparent that the graphs of $p_{2}, p_{3}, \ldots, \ldots$ 
converge to
the graph of $p_{\infty}$. This statement can be
proved rigorously :

\begin{theorem}\label{annealed}
For every $\hat{c}>0$, $\lim_{n\goesto
\infty}p_{n}(\hat{c})=p_{\infty}(\hat{c})$. 
\end{theorem}
\vspace{5mm}

\begin{figure}
\centerline{
\epsfig{figure=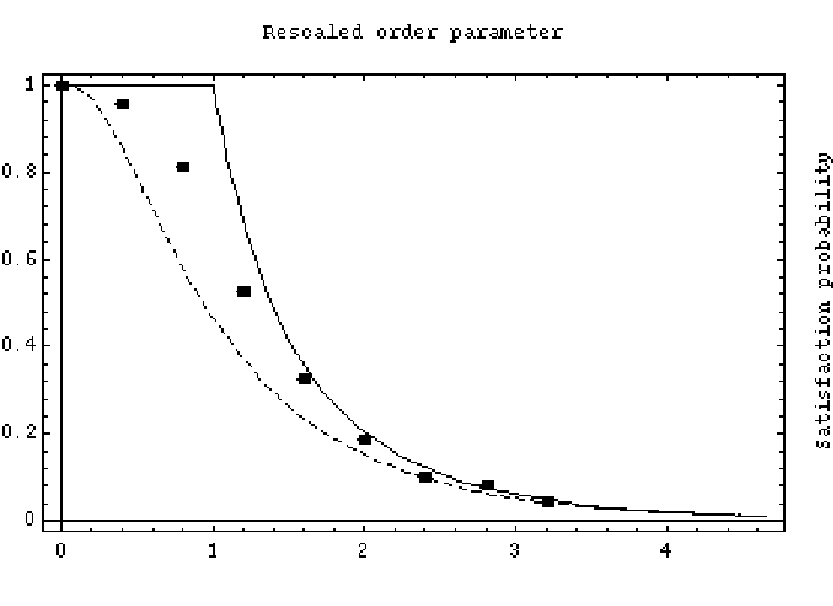,width=3.5in}}
\caption{Rescaled threshold functions}
 \end{figure}

As a bonus our analysis yields the following result:
\vspace{5mm}
\begin{theorem}\label{k:2:runtime}
Let $q$ be the limit of the 
expected number of iterations of \PUR\ on a random formula
$\Phi \in \Omega(2,n,cn)$, conditional on $\Phi$ being
satisfiable. Then 
\begin{equation}
\label{q:2}
q=
\left \{\begin{array}{ll}
                 \frac{1}{1-p_{2}\lambda_{2}c} & \mbox{, if $c\neq 
\frac{3}{2}$,}
                 \\
 
                 \infty,  & \mbox{ otherwise.}\\ 
        \end{array}
\right.
\end{equation} 
\end{theorem}
\vspace{5mm}

This theorem suggests (see Fig.2) and explains the ``easy-hard-easy''
pattern for the average running time of
\PUR\ 
in this case. Experiments we performed confirm this prediction.
 \begin{figure}
 \centerline{
 \epsfig{figure=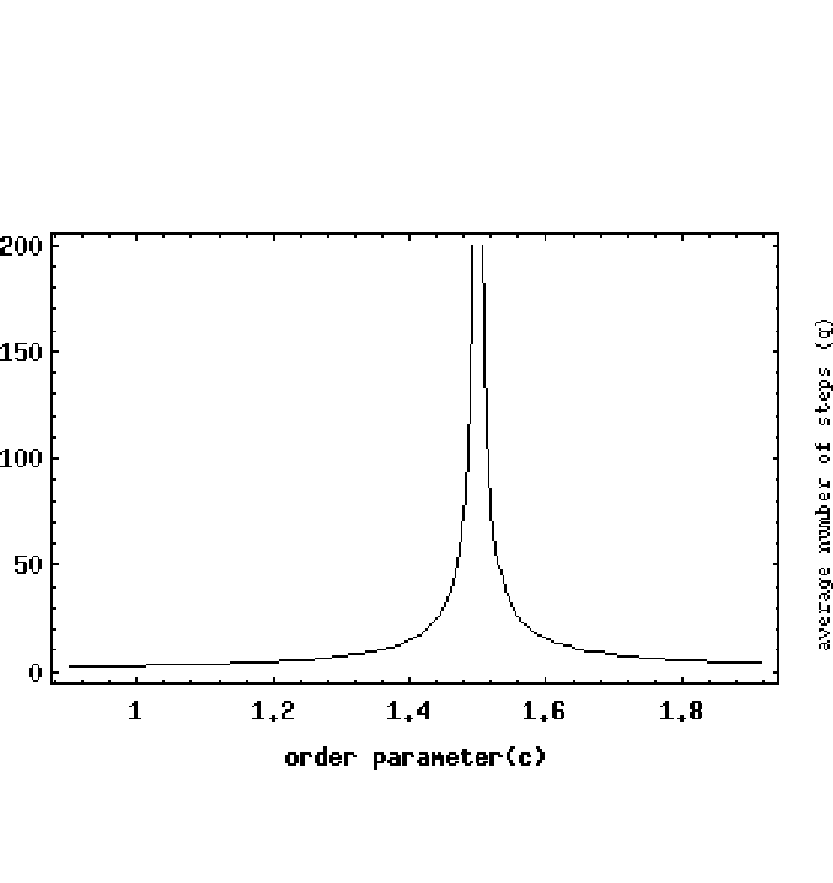,width=3.5in}}
 \caption{The ``easy-hard-easy'' pattern.}
 \label{figure-2}
 \end{figure}

\section{Preliminaries}

Throughout this paper we use ``with high probability'' (w.h.p.)
as a substitute for ``with probability $1-o(1)$''. 
We denote (sometimes abusing notation) by $B(n,p) (Po(\lambda))$ a 
random
variable having a binomial (Poisson) distribution with the 
corresponding
parameter(s), and by $a\cminus b$ the value $max(a-b,0)$. 
We will use the following version of the Chernoff bound
\vspace{5mm}

\begin{theorem}
If $0<\theta <1/4$ then
$\PR[|B(n,p)-np|>\theta np ] \leq e^{-np\frac{\theta^{2}}{4}}$.
\end{theorem}
\vspace{5mm}

as well as the related inequality from \cite{probabilistic-method} : 
 \vspace{5mm}

\begin{proposition}\label{chernoff:poisson}
Let $P$ have Poisson distribution with mean $\mu$. For $\epsilon >0$,
 
\[ \Pr[P\leq \mu \cdot (1-\epsilon)] \leq e^{\epsilon^{2}\cdot \mu
/2},
\]
 
\[ \Pr[P\geq \mu \cdot (1+\epsilon)] \leq
[e^{\epsilon}(1+\epsilon)^{-(1+\epsilon)}]^{\mu}.
\]
\end{proposition}    
\vspace{5mm}

We also use the following inequality:
\vspace{5mm}

\begin{proposition}
Let $k\in \N$ and $p\in [0,1]$. Then for every $n\geq k$ 
\begin{equation}
1-\sum_{i=0}^{k-1} {{n}\choose {i}}p^{i}(1-p)^{n-i}\leq {{n}\choose 
{k}}p^{k}.
\end{equation} 
\end{proposition}
\vspace{5mm}

\begin{PROOF} Define $f:[0,1]\goesto R$, $f(p)=1-\sum_{i=0}^{k-1} 
{{n}\choose
{i}}p^{i}(1-p)^{n-i} -{{n}\choose {k}}p^{k}$. It is easy to see that 
$f^{\prime}(p)=n{{n-1}\choose {k-1}}p^{k-1}[(1-p)^{n-k}-1]\leq 0$,
therefore $f$ is monotonically decreasing, and $f(0)=0$.
\end{PROOF}

We will also employ {\em couplings of Markov
chains} (see \cite{lindvall:coupling}) to assert stochastic
domination. The following is the definition of the type of 
coupling we employ in
this paper:
\vspace{5mm}
\begin{definition}
Let $(X_{t})_{t}$ and $(Y_{t})_{t}$ be two Markov chains on ${\bf Z}$. 
A {\em coupling of $X$ and $Y$ such that $X_{t}\leq Y_{t}$} is a
Markov chain $Z=(Z_{t,1},Z_{t,2})$ such that:
\begin{itemize}
\item $Z_{t,1}$ is distributed like $X_{t}$ given $X_{0}$. 
\item $Z_{t,2}$ is distributed like $Y_{t}$ given $Y_{0}$.
\item for every $i\geq 0$, $Z_{i,1}\leq Z_{i,2}$.
\end{itemize} 
\end{definition}
\vspace{5mm}

We use such couplings to bound the probability that a Markov 
chain $Y_{t}$ ever decreases below a certain value $a$ by coupling it 
with a chain $X_{t}$ such that $X_{t}\leq Y_{t}$ and using the
estimate $\Pr[\exists t: Y_{t}\leq a]\leq \Pr[\exists t: X_{t}\leq a]$ 
(that follows from the coupling). The couplings we construct employ the 
following ideas:   
\begin{itemize}
\item Suppose the recurrences 
describing $\Delta X_{t}$ and $\Delta Y_{t}$ are identical, 
except for one term, which is $B(m_{1},\tau)$ in $X_{t}$ and 
$B(m_{2},\tau)$ in $Y_{t}$, 
where $m_{1}\leq m_{2}$ are positive integers and $\tau \in (0,1)$. 
Obtain a coupling by identifying $B(m_{1},\tau)$ with the outcome of 
the first $m_{1}$ Bernoulli experiments in $B(m_{2},\tau)$. 
\item Suppose now that $\Delta X_{t}$ and $\Delta Y_{t}$ differ by
exactly one term which is $B(m,p)$ in $\Delta X_{t}$ and  
$B(m,q)$ in $\Delta Y_{t}$, $p \leq q$. Let $A_{i}$ and $B_{i}$,
$i=1,m$,  
be independent $0/1$ experiments with success probabilities $p$
and $\frac{q-p}{1-p}$ respectively. Define the pair $(Z_{t,1}, 
Z_{t,2})$ so that 
\begin{enumerate}
\item $Z_{t,1}$ is the number of times $A_{i}$ succeeds. 
\item $Z_{t,2}$ is the number of times at least one of $A_{i}$ and 
$B_{i}$
succeeds. 
\end{enumerate}
\end{itemize}


We measure the distance between two probability distributions 
$P$ and $Q$ by {\em the total variation distance}, 
denoted by $d_{TV}(P,Q)$,  and recall the following results, 
(see \cite{sheu:poisson} and \cite{barbour:holst:janson}, page
2 and Remark 1.4):
\vspace{5mm}

\begin{lemma}\label{b:h:j}If $n,p,\lambda, \mu >0$ then 
$d_{TV}(B(n,p),Po(np))\leq \min\{np^{2},\frac{3p}{2}\}$ and
$d_{TV}(Po(\lambda), Po(\mu))\leq |\mu - \lambda|$. 
\end{lemma}
\vspace{5mm}

We will also need the following simple lemma:

\begin{lemma}\label{approximation}
Let c be a fixed positive integer. For every $t\in \N$ let
$\xi_{t}$, $\eta_{t}$ be two probability distributions. Define the
Markov chains $(X_{t})_{t}$ and $(Y_{t})_{t}$ by recurrences 
\begin{equation}
\left\{\begin{array}{l}
X_{t+1}=X_{t}\cminus c + \xi_{t}, \\
Y_{t+1}=Y_{t}\cminus c + \eta_{t}.\\
\end{array}
\right.
\end{equation}

Then, for every $t\geq 0$, $d_{TV}(X_{t},Y_{t})\leq
d_{TV}(X_{0},Y_{0})+ \sum_{i=0}^{t-1} d_{TV}(\xi_{i}, \eta_{i}).$  
\end{lemma}

\beginproof 

The following result gives a more convenient inequality that
immediately implies Lemma~\ref{approximation}
\vspace{5mm}

\begin{lemma}\label{easy:approximation}
Let c be a fixed positive integer. Let
$X$, $Y$, $\xi$, $\eta$ be random variables with nonnegative integer
values. Define the 
random variables $Z$ and $T$ by recurrences 
\begin{equation}
\left\{\begin{array}{l}
Z=X\cminus c + \xi, \\
T=Y\cminus c + \eta.\\
\end{array}
\right.
\end{equation}
Then, for every $d_{TV}(Z,T)\leq
d_{TV}(X,Y)+ d_{TV}(\xi, \eta).$  
\end{lemma}
\vspace{5mm}

\beginproof

To prove this result, we will denote (for the ``generic'' r.v. $A$) by
$A_{i}$ the probability that $A$ takes value $i$. We also employ the 
following simple inequality, valid for $a,b,c,d\geq 0$: $|ad-bc|\leq
a|d-c|+|a-b|c$. 

For every $a\geq 0$ we have:
\[
Z_{a}=\sum_{i=0}^{c} X_{i}\xi_{a}+\sum_{i=c+1}^{c+a} X_{i}\xi_{a+c-i},
\]
\[
T_{a}=\sum_{i=0}^{c} Y_{i}\eta_{a}+\sum_{i=c+1}^{c+a} 
Y_{i}\eta_{a+c-i},
\]

Applying the above-mentioned inequality and summing we get:

\begin{eqnarray*}
d_{TV}(Z,T) \\ & \leq &
\frac{1}{2} \{
\sum_{i=0}^{c}\sum_{a=0}^{\infty}X_{i}|\xi_{a}-\eta_{a}|
+\sum_{i=0}^{c}\sum_{a=0}^{\infty}|X_{i}-Y_{i}|\eta_{a}+ \\
& + & 
\sum_{i=c+1}^{c+a}\sum_{a=0}^{\infty}X_{i}|\xi_{c+a-i}-\eta_{c+a-i}|
+\sum_{i=c+1}^{c+a}\sum_{a=0}^{\infty}|X_{i}-Y_{i}|\eta_{c+a-i}\}.
\end{eqnarray*}

Let A,B,C,D be the four terms of the sum. By simple algebraic
manipulations we obtain:
\[
\begin{array}{lcl}
 A = (\sum_{i=0}^{c}X_{i})\cdot d_{TV}(\xi,\eta), &\hspace{5mm} & B = 
\frac{1}{2}\sum_{i=0}^{c} |X_{i}-Y_{i}|,\\ 
C = (\sum_{i=c+1}^{\infty}X_{i})\cdot d_{TV}(\xi,\eta),
& \hspace{5mm} & D =
 \frac{1}{2}\sum_{i=c+1}^{\infty}|X_{i}-Y_{i}|,
\end{array}
\]
and the result follows.
\qed

Finally, we need the following trivial occupancy property:
\vspace{5mm}

\begin{lemma}\label{occupancy}
Let $a$ white balls and $b$ black balls be thrown uniformly at random
in $n$ bins. 
\begin{enumerate}
\item if $r=\max(a,b)=o(n^{1/2})$ then the probability that there is a 
bin that contains both white and black balls is at most 
$\frac{4r^2}{n}=o(1)$.
\item if $s=\min(a,b)=\omega(n^{1/2})$ then the probability that there 
is a 
bin that contains both white and black balls is $1-o(1/poly)$.
\end{enumerate}
\end{lemma}
\vspace{5mm}

\beginproof
The first part is easy: the probability that two balls (of any color) 
end up in the same bin is at most ${{a+b}\choose {2}}\cdot 
\frac{1}{n}$.
For the second part, let $A$ be the event that no two balls of
different colors end up in the same bin, and let $B$ the event that at
least $\sqrt{n}$ bins contain white balls. We have:
\[ \Pr[A]\leq \Pr[A|B]+\Pr[\overline{B}].\]
But
\[ \Pr[\overline{B}]\leq {{n}\choose {\sqrt{n}}}\cdot
(\frac{1}{\sqrt{n}})^{a}= n^{\sqrt{n}-a/2}=o(\frac{1}{poly}), \]
and 
\[\Pr[A|B]\leq (1-\frac{1}{\sqrt{n}})^{b}\sim
e^{-b/\sqrt{n}}=o(\frac{1}{poly}). \]
\qed

The algorithm \PUR\ is displayed in Figure 3. 
We regard \PUR\ as working in stages, indexed by the
number of variables still left unassigned; thus, the stage number
decreases as \PUR\ moves on. We say that {\em formula $\Phi$ survives
Stage $t$} if \PUR\ on input $\Phi$ does not halt at Stage $t$ or
earlier. Let $\Phi_i$ be the formula at the
beginning of stage $i$, and let $N_{i}$ denote the number of its
clauses. We will also denote by $P_{i,t} (N_{i,t})$, the number of 
clauses of
$\Phi_{t}$ of size $i$ and containing one (no) positive
literal. Define $\Phi_{i,t}^{P}$ ($\Phi_{i,t}^{N}$) to be the
subformula of $\Phi_{t}$ containing the clauses counted by $P_{i,t} 
(N_{i,t})$.

The following lemmas were proved in \cite{istrate:cs.DS/9912001}, in 
the
context of analyzing the behavior of \PUR\ on $\Phi\in
\Omega(n,n,m)$, $m=c\cdot 2^n$.
\vspace{5mm}

\begin{lemma}\label{k:inf:recurrence}
\begin{enumerate}
\item 
Suppose $\PUR$ does not halt before stage $t$. Then, conditional on $N_{t}$,
the clauses of $\Phi_{t}$ are random and independent. 
\item   
Suppose now that we condition on $\Gamma_{t}=(N_{1,t},N_{2,t},P_{1,t},
P_{2,t}$ and on the fact that $\Phi$
survives Stage $t$ as well. Then  we have 

\begin{equation}\label{eq:markovchain}
N_{t-1}=N_{t}-\Delta_{1,P}(t)-\Delta_{2,P}(t), 
\end{equation}

where 
\begin{itemize}
\item $\Delta_{1,P}(t)$, the number of positive clauses that are
satisfied at stage $t$, has the distribution $1+B\left(P_{1,t}-1,\frac{1}{t}\right)$. 
\item  
$\Delta_{2,P}(t)$, the number of positive non-unit clauses 
that are satisfied at stage $t$, has the binomial distribution
$B\left(P_{2,t},\frac{1}{t}\right)$.
\end{itemize}
\end{enumerate}
\end{lemma}

\vspace{5mm}

\begin{lemma}\label{k:inf:bounds}
For every $c>0$ and every $t, n-c\sqrt n \leq t \leq n$, 
the conditional probability that the inequality 
\begin{equation}\label{concentrate}
N_{n}-(n-t)\left[1+\frac{2(N_{n}-1)}{t}\right]\leq N_{j}\leq 
N_{n}\end{equation}
holds for all $t\leq j \leq n$, in the event that $\PUR$ reaches stage 
$t$,
is $1-o(1)$.
\end{lemma}
\vspace{5mm}

\begin{lemma}\label{k:inf:prob}
Let $X_{n}\in [0,n]$ be the r.v. denoting the number of iterations of 
\PUR\ on
a random {\em satisfiable} formula $\Phi\in \Omega(n,c\cdot
2^{n})$. Then $X_{n}$ converges in distribution to a distribution 
$\rho$ on $[0,n]$ having support on the nonnegative integers, 
$\rho=(\rho_{k})_{k\geq
0}$, $\rho_{k}= Prob[\rho = k]$, 
given by
\[ \rho_{k}=\frac{e^{-2^{k}c}}{1-F(e^{-c})}\cdot \prod_{i=1}^{k-1}
(1-e^{-2^{i}c}).
\]
\end{lemma}
\vspace{5mm}

\begin{center}
\begin{figure}
{\tt
\begin{tabbing}

Pr\=ogram PUR($\Phi$): \\
   \> if \= $\Phi$ (contains no positive literal as a clause)\\
   \> \>then \= return TRUE \\
   \> \>else \\
   \> \> \>choose such a positive unit clause $x$ \\
   \> \> \>if \= ($\Phi$ contains $\overline{x}$ as a clause)\\
   \> \> \> \>then \= \\
   \> \> \> \> \>return FALSE \\
   \> \> \> \>else \\
   \> \> \> \> \>let $\Phi^{\prime}$ be the formula \\
   \> \> \> \> \>obtained by setting
   $x$ to 1 \\
    \> \> \> \> \>return \PUR($\Phi^{'}$) \\
\end{tabbing}
}
\caption{Algorithm PUR}
\end{figure}
\end{center}
\section{The proof of Theorem~\ref{k:infinite}}

Let $c_{1}<c_{2}<c_{3}$ be arbitrary constants. Consider three
 random formulas $\Phi_{1}\in \Omega(n,{\bf k(n)},c_{1}\cdot
 \frac{H_{k(n)}}{n})$,$\Phi_{2} \in \Omega(n,{\bf n},c_{2}\cdot
 2^{n})$ and  $\Phi_{3}\in\Omega(n,{\bf k(n)}, c_{3}\cdot 
\frac{H_{k(n)}}{n})$,
and let $\Phi^{\prime}$ be the subformula of $\Phi_{2}$ consisting of
 the clauses of size at
most $k(n)$. By the Chernoff bound, with high probability,
$m^{\prime}$, the number of clauses of $\Phi^{\prime}$, is in the 
interval
$[c_{1}\cdot \frac{H_{k(n)}}{n},c_{3}\cdot \frac{H_{k(n)}}{n}] $. 
When $n\goesto \infty$ the probability that $\Phi_{2} \in \HSAT$ tends
to $1-F_{1}(e^{-c_{2}})$. 

From Lemma~\ref{k:inf:prob} we infer the following easy consequence
\vspace{5mm}
\begin{claim}
The probability that \PUR\  accepts $\Phi_{2}$ 
after stage $n-k(n)+1$ is $o(1)$. 
\end{claim}

Since in the first $k(n)-1$ stages 
of \PUR\  {\em only the clauses of $\Phi^{\prime}$ can influence the
algorithm acceptance/rejection of  $\Phi_{2}$ 
(because \PUR\  accepts/rejects
at Stage $i$ based only on the unit clauses, and 
each non-simplified clause loses at most one literal at each phase)},
\[ |\Pr[\Phi_{2}\in \HSAT]- \Pr[\Phi^{\prime}\in \HSAT]|= o(1).
\]
By the monotonicity of SAT and the randomness of
$\Phi_{1},\Phi_{2}, \Phi^{'}$ we have 

\[ \Pr[\Phi_{1}\in \HSAT]-o(1)\leq \Pr [\Phi_{2} \in \HSAT] \leq
\Pr[\Phi_{3}\in \HSAT]+o(1). 
\]
Taking limits it follows that 

\begin{eqnarray*}
{\overline{\lim}_{n\goesto \infty} \Pr}_{\Phi\in
\Omega(n,k(n),c_{1}H_{k(n)}/n)} [\Phi \in \HSAT] & \leq 1-F(e^{-c_2}) 
\leq & \\
{\underline{\lim}_{n\goesto \infty} \Pr}_{\Phi \in
\Omega(n,k(n),c_{3}H_{k(n)}/n)} [\Phi \in \HSAT] .
\end{eqnarray*}
Since $c_{1},c_{2},c_{3}$ were chosen arbitrarily, 
by choosing $c_{1}=c, c_{2}=c+\epsilon$, and $c_{2}=c-\epsilon, c_{3}=
c$, respectively, we infer that 

\begin{eqnarray*}
1-F_{1}(e^{-(c-\epsilon)})\leq  {\underline{\lim}_{n\goesto
\infty} \Pr}_{\Phi\in \Omega(n,k(n),cH_{k(n)}/n)}[\Phi \in
\HSAT]  \leq & \\ 
{\overline{\lim}_{n\goesto \infty}\Pr}_{\Phi \in
\Omega(n,k(n),cH_{k(n)}/n)}[\Phi \in \HSAT]\leq 
1-F_{1}(e^{-(c+\epsilon)}).
\end{eqnarray*}
As $\epsilon$ is arbitrary, we get the desired result.
\qed

\begin{observation}\label{obs:coupling}
One point about the previous proof that is intuitively clear, but gets
somewhat obscured by the technical details of the proof, is that if
$\Phi_{2} \in \Omega(n,{\bf n},c_{2}\cdot 2^{n})$
then $\Phi^{'}$ behaves ``for every practical purpose'' 
as if it were a uniform formula in $\Omega(n,{\bf k(n)},c_{2}\cdot
 \frac{H_{k(n)}}{n})$. We will use a similar
intuition in the proof of Proposition~\ref{annealed}. 
\end{observation}
 \vspace{5mm}

\section{The uniformity lemma}

The following lemma is the analog of Lemma~\ref{k:inf:recurrence}
for the case $k=2$, and the basis for our analysis of this case:
\vspace{5mm}

\begin{lemma}\label{k:2:recurrence}
Suppose that $\Phi$ survives up to stage $t$. Then, conditional on 
$(P_{1,t}, N_{1,t}, P_{2,t}, N_{2,t})$, the clauses in  
$\Phi_{1,t}^{P},
\Phi_{1,t}^{N}, \Phi_{2,t}^{P}, \Phi_{2,t}^{N}$ are chosen uniformly
at random and are independent. Also, conditional on the
fact that $\Phi$ survives stage $t$ as well, the following recurrences
hold:
\begin{equation}\label{k:2:markovchain}
\left \{\begin{array}{l}
         P_{1,t-1}=P_{1,t}-1-\Delta_{1,t}^{P}+\Delta_{12,t}^{P}, \\        
         N_{1,t-1}=N_{1,t}+\Delta_{12,t}^{N},                    \\
         P_{2,t-1}=P_{2,t}-\Delta_{12,t}^{P}-\Delta_{02,t}^{P},  \\
         N_{2,t-1}=N_{2,t}-\Delta_{12,t}^{N},                    \\
        \end{array}
\right.
\end{equation}
where (in distribution)
\begin{equation}\label{k:2:distribution}
\left \{\begin{array}{l}
\Delta_{1,t}^{P} =B(P_{1,t}-1,1/t),\\
\Delta_{12,t}^{P}=B(P_{2,t},1/t),\\
\Delta_{02,t}^{P}=B(P_{2,t}-\Delta_{12,t}^{P},1/t),\\
\Delta_{12,t}^{N}=B(N_{2,t},2/t).\\
\end{array}
\right.
\end{equation}
\end{lemma}
 \vspace{5mm}

\beginproof
A formula will be represented by an
$m\times 2$ table. The rows
of the table correspond to clauses in the formula and the entries are
its literals. They are gradually unveiled as the algorithm proceeds. 
We assume that when generating $\Phi$ we mark those
clauses containing only one literal (so that we know their location,
but not their content).
We say that a row (or a clause) is ``blocked'' either if the clause is
already satisfied or the clause has been turned into the empty
clause.
Suppose $\PUR$ arrives at stage $t$ on $\Phi$.  Then in stages
$i=n, n-1, \ldots, t+1$, $\Phi_i$ should contain a unit clause
consisting of a positive literal but should not have contained
complementary unit clauses of the same variable.
To carry out the disclosure at stage $i$, let $x$ be the variable set
to one in this stage. We assume that the formula unveils
all occurrences of $x$ or $\overline{x}$ in $\Phi$. For each clause we 
perform the following:

\begin{enumerate}
\item if it contains $x$ we unveil all its literals and block;
\item otherwise we do nothing. 
\end{enumerate}
The clauses of $\Phi_{t}$ having size two correspond to the rows of 
$\Phi$
that contain no unveiled literal. 
The clauses of size one are either the clauses of
size one in $\Phi$ that contain none of the chosen literals, or the 
clauses of size two that contain the negation of one chosen variable 
and another is yet to be chosen. 
Given these observations the uniformity and independence follow from
the way we construct $\Phi$. 

To prove the recurrences, let $x$ be the variable set to
one in stage $t$ (it exists since \PUR\ does not halt at
this stage). By uniformity and independence, each of the $P_{1,t}-1$
positive unit clauses of $\Phi_{t}$, other than the chosen one, is
equal to $x$ with probability $1/t$ (since there are $t$ variables
left at this stage). On the other hand, the positive unit clauses of 
$\Phi_{t-1}$ that are not present
in $\Phi_{t}$ can only come from clauses of size two of $\Phi_{t}$
that contain $\overline{x}$ and a positive literal (therefore counted
by $P_{2,t}$). Uniformity and independence imply therefore that
$\Delta_{1}^{P}(t)$ has the distribution claimed in
(\ref{k:2:distribution}). The other relations can be
justified similarly (noting that, since \PUR\ does not reject at this
stage, every negative unit clause of $\Phi_{t}$ is also present in 
$\Phi_{t-1}$).

It will be useful to consider the Markov chain
(\ref{k:2:markovchain}) for all
values of $t=n,\ldots, 0$ (even when the algorithm halts). To
accomplish that, the ``minus'' signs in the first equation of
(\ref{k:2:markovchain}) and the definition of $\Delta_{1,t}^{P}$ 
should be replaced by $\cminus$. We also need to specify the
distribution of each component of
the tuple $(P_{1,n}, N_{1,n}, P_{2,n}, N_{2,n})$. Let $\Delta_{n}$ be
a random variable having the Bernoulli distribution $B(cn,
\frac{2n}{2n+3{{n}\choose {2}}})$. It is easy to see that in
distribution
 
\begin{equation}\label{k:2:initial:condition}
\left \{\begin{array}{l}
         P_{1,n}=B(\Delta_{n},1/2),\\
         N_{1,n}=\Delta_{n}-P_{1,n},\\
         P_{2,n}=B(cn-\Delta_{n},2/3)\\
         N_{2,n}=cn-\Delta_{n}-P_{2,n}.\\
        \end{array}
\right.
\end{equation}
\endproof
\qed 

\section{Proof of Theorem~\ref{k:2}}

The main intuition for the proof is that in ``most interesting stages''
$\Delta_{1,t}^{P}=0$ and $\Delta_{12,t}^{P}$ is approximately
 Poisson distributed. Therefore,  $P_{1,t}$ qualitatively
behaves like the Markov Chain $(Q_{t})_{t}$ defined by 
\begin{equation}
\left \{\begin{array}{l}
        Q_{n+1}=1,\\
        Q_{t-1}=Q_{t}\cminus 1+Po(\lambda),\\
\end{array}
\right.
\end{equation}
where $\lambda=2c/3$.
This explains the closed form of the limit probability: a well-known 
result 
states that $\rho$, the probability that the queuing chain $Q_{t}$ 
reaches 
state 0, satisfies the equation $\rho= \Phi(\rho)$, where
$\Phi(t)=e^{\lambda(t-1)}$ is the generating function of the
arrival distribution $Po(\lambda)$.  
We will define a suitable value $\omega_{0}$ such that:
\begin{enumerate}
\item With high probability \PUR\ does not reject in any of stages $n,
\ldots, n-\omega_{0}$. 
\item \PUR\ accepts ``mostly before or at stage $n-\omega_{0}$'' (i.e. 
the 
probability that \PUR\ accepts after stage $n-\omega_{0}$, given that
$\Phi$ survives this far is $o(1)$). 
\item With high probability, for every $t\in n, \ldots, n-\omega_{0}$, 
$\Delta_{1,t}^{P}=0$.  
\item At stages $n,\ldots, n-\omega_{0}$, $P_{1,t}$ is ``very close''
to $Q_{t}$, with respect to total variation distance. 
\end{enumerate}
 
This program can be accomplished as described if $c< 3/2$. To prove
Property 4 we make use of Lemmas~\ref{b:h:j} and
\ref{occupancy}. Property 2 is proved only implicitly: in this
case (see \cite{hoel:port:stone}) the probability that $Q_{i}=0$ for
some $i$ tends to one, and, in fact, by a technical result due to
Frieze and Suen (Lemma 3.1 in \cite{frieze-suen}), $\Pr[Q_{i}=0\mbox{ 
for some
}i\geq n - \log n]$ is $1-o(1)$.

Let us now concentrate on the case when $c>3/2$ (the case when $c=3/2$ 
will
follow by a monotonicity argument). In the previous argument we only 
used 
the fact
that $c<3/2$ when deriving the probability that $Q_{t}$ hits state 0,
hence the
arguments from above carry on, and the conclusion is that the
probability that \PUR\ accepts at one of the stages $n,\ldots,
n-\omega_{0}$ differs by $o(1)$ from the probability that $Q_{t}=0$
somewhere in this range. We now, however, have to consider the
probability that \PUR\ accepts at some stage later than $n-\omega_{0}$
and aim to prove that this probability is $o(1)$. It is conceptually 
simpler to divide
the interval $[n-\omega_{0},0]$ into two subintervals, $[n-\omega_{0},
n-\omega_{1}]$ and its complement, such that
w.h.p. $\Phi_{n-\omega_{1}}$ (if defined) contains two opposite unit
clauses, therefore
the probability that \PUR\ accepts after stage $n-\omega_{1}$ is
$o(1)$. In the range $[n-\omega_{0},n-\omega_{1}]$ we would like to
prove that ``most of the time'' $\Delta_{1,t}^{P}$ is zero and
$P_{1,t}$ is ``close'' to $Q_{t}$ and to reduce the problem to the
analysis of $Q_{t}$. Unfortunately there are two problems with this
approach: although the probability that each individual
$\Delta_{1,t}^{P}>0$ is fairly small, to make $\Phi_{n-\omega_{1}}$
unsatisfiable w.h.p., $\omega_{1}$ has to
be $\omega(\sqrt n)$. This implies
that we cannot sum these probabilities over
$[n-\omega_{0},n-\omega_{1}]$ and expect the sum to be $o(1)$; a
similar problem arises if we want to sum the upper bounds for
$d_{TV}(\Delta_{12,t}^{P},Po(\lambda))$.

Fortunately there is a way to circumvent this, avoiding the use
of total variation distance altogether: although we cannot guarantee 
that
w.h.p. each $\Delta_{1,t}^{P}=0$, we can arrange that w.h.p. for every
sequence of $p$ consecutive stages $t, t-1, \ldots t-p+1$, 
$\Delta_{1,t}^{P}+\Delta_{1,t-1}^{P}+\ldots +\Delta_{1,t-p+1}^{P}\leq
3$ (*). Intuitively, in any sequence of $p$ consecutive steps at most
$p+3$
clients leave the queue, and the number of those who arrive is the sum
of $p$ approximately Poisson variables, thus approximately Poisson
with parameter $p\lambda$. Choosing $p$ large enough so that $\lambda
>1+\frac{3}{p}$ ensures that in any $p$ steps {\em the average number 
of  
customers that arrive is strictly larger than the number of customers
that are served in this time span}. Therefore we will seek to 
approximate
$P_{1,t}$ by a queuing chain $\overline{Q}_{t}$ with this
property. Since $P_{1,n-\omega_{0}}=\overline{Q}_{n-\omega_{0}}$ is
``large,'' an elementary analysis of the queuing chain implies 
that the probability that 
$\overline{Q}_{t}$ hits state 0 in the interval
$[n-\omega_{0},n-\omega_{1}]$ is exponentially small. So we obtain the
desired result if $\overline{Q}_{t}$ is constructed so that it is
stochastically dominated by $P_{1,t}$. 
 
\subsection{The case $c<3/2$} 
  Define $\omega_{0}=n^{0.1}$.
The following are the main steps of the proof in this case:
\vspace{5mm}

\begin{lemma}\label{small:p2}  
With probability $1-o(1/poly)$ for every $t\in [n,\ldots , n/2]$ we
have  $$\Delta_{12,t}^{P},\Delta_{02,t}^{P}, \Delta_{12,t}^{N}\leq
\frac{1}{2}n^{0.1}.$$ 
\end{lemma}
\vspace{5mm}

\beginproof
Use the coupling with $m_{1}=P_{2,t} (N_{2,t})$,
$m_{2}=cn$, $\tau = 1/t$, and apply Chernoff bound to
$B(cn,1/t)$.
\endproof
\qed
\vspace{5mm}

\begin{corrolary}\label{p:2:t}  
Consider $\omega \leq n/2$. 
If for every $t\in [n,\ldots , n/2]$,
$\Delta_{12,t}^{P},\Delta_{02,t}^{P}, \Delta_{12,t}^{N}\leq 
\frac{1}{2}n^{0.1}$ then, for all $t\in
[n,\ldots, n-\omega]$, $P_{1,t}, N_{1,t}, |P_{2,t}-P_{2,n}|, 
|N_{2,t}-N_{2,n}| <
(n-t)\cdot n^{0.1}$. 
\end{corrolary}
\vspace{5mm}

\begin{lemma}\label{small:delta1}
If for all $t\in
[n,\ldots, n-\omega]$, $P_{1,t}, N_{1,t}, |P_{2,t}-P_{2,n}|, 
|N_{2,t}-N_{2,n}| <
(n-t)\cdot n^{0.1}$ holds then 
w.h.p. $\Delta_{1,t}^{P}=0$ for every $t\in [n,\ldots , n-\omega_{0}]$. 
\end{lemma}
\vspace{5mm}

\beginproof
$\Pr[B(P_{1,t}-1,\frac{1}{t})>0] = 1-\Pr[B(P_{1,t}-1,\frac{1}{t})=0]= 
1-(1-\frac{1}{t})^{P_{1,t}-1}<\frac{P_{1,t}-1}{t}< n^{-0.9}$.  
\endproof
\qed
\vspace{5mm}

\begin{lemma}\label{p:2:n}
W.h.p., $|P_{2,n}-\frac{2}{3}cn|, |N_{2,n}-\frac{1}{3}cn|  < n^{0.6}$. 
\end{lemma}
\vspace{5mm}

\beginproof
Directly from the Chernoff bounds on $\Delta_{n}$ and $P_{2,n}$.
\endproof
\qed
\vspace{5mm}

\begin{lemma}\label{p:2:t:poisson}
If the events in the conclusions of Lemmas~\ref{p:2:t} and \ref{p:2:n} 
hold for 
$\omega = \omega_{0}$, $\epsilon_{1}=1/6$ and $\epsilon_{2}=0.1$, then 
there exists a constant
$r>0$ such that for every $t=n, \ldots,n-\omega_{0}$,
$|\frac{P_{2,t}}{t}-\frac{2}{3}c| \leq r n^{-0.4}$. 
\end{lemma}
\vspace{5mm}

\beginproof
 We have
$$|\frac{P_{2,t}}{t}-\frac{2}{3}c| \leq
P_{2,t}\left|\frac{1}{t}-\frac{1}{n}\right|+
\frac{|P_{2,t}-P_{2,n}|}{n}+ 
\left|\frac{P_{2,n}}{n}-\frac{2}{3}c\right | \leq
P_{2,n}\frac{\omega_{0}}{n(n-\omega_{0})}+\frac{n^{0.2}}{n}+ 
n^{0.6-1},$$
by Lemma ~\ref{p:2:t:poisson} and $n-\omega_{0}\leq t\leq n,$ and the 
result
immediately follows.
\endproof
\qed
\vspace{5mm}

\begin{lemma}\label{distance}
If the conclusions of Lemmas~\ref{p:2:t:poisson} and
\ref{small:delta1} are true  then 
$$\sum_{t=n-\omega_{0}}^{n}d_{TV}(P_{1,t},Q_{t})=o(1/\omega_{0}).$$
\end{lemma}
\vspace{5mm}

\beginproof
By
Lemma~\ref{p:2:t:poisson} and the inequalities on
total variation distance there exist $r_{1},r_{2}>0$ such that 
\begin{eqnarray*}
d_{TV}(\Delta_{12,t}^{P},Po(\lambda)) & \leq & 
d_{TV}\left(\Delta_{12,t}^{P}, Po\left(\frac{P_{2,t}}{t}\right)\right)+
d_{TV}\left(Po\left(\frac{P_{2,t}}{t}\right) 
,Po\left(\frac{2}{3}c\right)\right) \\ & \leq & r_{1}\frac{1}{t}+r_{2}n^{-0.4}\leq 
r_{3}n^{-0.4}, 
\end{eqnarray*} where $r_{3}=r_{1}+r_{2}$. Employing
Lemma~\ref{approximation} it follows that
$$\sum_{t=n-\omega_{0}}^{n}d_{TV}(P_{1,t},Q_{t})\leq
r_{3}\sum_{t=n-\omega_{0}}^{n}tn^{-0.4}\leq
r_{3}n^{-0.4}\frac{\omega_{0}^{2}}{2},$$ and this amount is 
$o(1/\omega_{0})$. 
\endproof
\qed
\vspace{5mm}

\begin{observation}
The probability that the conditions in the previous lemma are not
fulfilled is at most $\omega_{0}^{4}/n = n^{-0.6}$. Indeed, 
the events that ensure the applicability of the previous lemma are: 
\begin{enumerate}
\item for every $t\in [n,\ldots , n/2]$, 
$\Delta_{12,t}^{P},\Delta_{02,t}^{P}, \Delta_{12,t}^{N}\leq
\frac{1}{2}n^{0.1}$,
\item for all $t\in
[n,\ldots, n-\omega_{0}]$, $\Delta_{1,t}^{P}=0$, and
\item  $|P_{2,n}-\frac{2}{3}cn|, |N_{2,n}-\frac{1}{3}cn|,  < n^{0.6}$
\end{enumerate}
The first and the third events have probability $1-o(1/poly)$ (as they
come from applying Chernoff bounds). The second fails (for a specific
$t$) with probability at most $\frac{P_{1,t}}{n-t}\leq 
\omega_{0}^{2}/(n-\omega_{0})$, so its total
probability is at most $\omega_{0}\cdot
\omega_{0}^{2}/(n-\omega_{0})$. Both terms can be absorbed into
$\omega_{0}^{4}/n$. 

\end{observation}
\vspace{5mm}

\begin{lemma}\label{no:reject}
If the event in Lemma~\ref{p:2:t} holds then 
w.h.p. \PUR\ does not reject at stage $t$, for every $t$ in the range 
$n$, $n-1, \ldots, n-\omega_{0}$, given that $\Phi$ survives up to
this stage.   
\end{lemma}
\vspace{5mm}

\beginproof
To prove Lemma~\ref{no:reject} we show that, 
with high probability the unit clauses of each $\Phi_{t}$ involve
different variables. This can be seen as follows: consider
$P_{1,t}+N_{1,t}$ balls to be thrown into $t$ urns. The probability
that two of them arrive in the same urn is at most
${{P_{1,t}+N_{1,t}}\choose {2}}\cdot \frac{1}{t}$. This is upper 
bounded
by $\frac{(\omega_{0}n^{0.1})^{2}}{2(n-\omega_{0})}$. Summing this for 
$t=n, \ldots,
n-\omega_{0}$ yields an upper bound, which is $o(1)$. 
\endproof
\qed
\vspace{5mm}

The proof for the case $c<3/2$ follows easily from these results: 
with probability $1-o(1)$ all the events in Lemmas~\ref{small:p2}, 
\ref{p:2:t}, \ref{small:delta1}, \ref{p:2:n}, \ref{distance}, and 
\ref{no:reject} take place, therefore 
\PUR\ does not reject at any of the stages $n$ to $n-\omega_{0}$ and 
$P_{1,t}$ is close to $Q_{t}$ in the sense of
Lemma~\ref{distance}. Therefore the probability that for some $t$ in
this range $P_{1,t}=0$ (i.e. \PUR\ accepts) differs by $o(1)$ from the
corresponding probability for $Q_{t}$. But according to the result by
Frieze and Suen \cite{frieze-suen} this latter probability is $1-o(1)$.
\endproof

\subsection{The case $c>3/2$}
Define $\omega_{1}=n^{0.51}$. The following
are the auxiliary results we use in this case:

\vspace{5mm}

\begin{lemma}\label{trick:delta} Let $A=n^{0.61}$. 
For every $k>0$ there exists a constant $c_{k}>0$ such that for every 
$r>0$ the probability that there exists $t\in [n-\omega_{0},
n-\omega_{1}]$, $\Delta_{1,t}^{P}+ \Delta_{1,t-1}^{P}+\ldots +
\Delta_{1,t-r+1}^{P}\geq k$ is at most 
$c_{k}(\omega_{1}-\omega_{0})(rA/n)^{k}$. 
\end{lemma}
\vspace{5mm}

\beginproof
By Corollary~\ref{p:2:t} we can assume that $P_{1,t}\leq A$. Then for 
every $i$, 

\[\Pr[\Delta_{1,t}^{P}\geq i]=\Pr[B(P_{1,t}-1,\frac{1}{t})\geq i]\leq
\Pr[B(A,\frac{1}{t})\geq i]
\]

\[ = 1 -
\sum_{j=1}^{i-1}{{A}\choose{j}}\left(\frac{1}{t}\right)^{j}\left(1-\frac{1}{t}\right)^{A-j}
\]

\[ \leq {{A}\choose{i}}(\frac{1}{t})^{i}
\]
The event $\Delta_{1,t}^{P}+ \Delta_{1,t-1}^{P}+\ldots +
\Delta_{1,t-r+1}^{P}\geq k$ happens when:
\begin{itemize}
\item one of the factors is at least $k$, or
\item one of the factors is at least $k-1$, and another one is at
least 1,  or
\item \ldots
\item at least $k$ of the factors are at least one. 
\end{itemize}
(a finite number of possibilities). Applying the previous inequality, 
and taking into account that $r,k$ are fixed immediately proves the 
lemma.
\endproof
\vspace{5mm}

To flesh out the argument outlined before we construct a
succession of Markov chains running along $P_{1,t}$, 
that provide better and better ``approximations'' to 
$\overline{Q}_{t}$. 
Our use of indices will be slightly nonstandard (to reflect
the connection with $P_{1,t}$), in that 
the sequence of indices starts with $n-\omega_{0}$ and is decreasing.
\vspace{5mm}
 \begin{definition}
Let
$X_{n-\omega_{0}}=Y_{n-\omega_{0}}=Z_{n-\omega_{0}}=\overline{Q}_{n-\omega_{0}}=
P_{1,n-\omega_{0}}$
and 
\begin{equation}\label{sequences}
\left \{\begin{array}{l}
         X_{t-1}=X_{t}-(p+3)\chi_{p{\bf 
Z}+1}(n-\omega_{0}-t)+\Delta_{12,t}^{P},
\\
         Y_{t-1}=Y_{t}-(p+3)\chi_{p{\bf Z}+1}(n-\omega_{0}-t)+B(P_{2, 
n-\omega_{
1}}, 1/t),\\
         Z_{t-1}=Z_{t}-(p+3)\chi_{p{\bf Z}+1}(n-\omega_{0}-t)+B(P_{2, 
n-\omega_{
1}},\frac{1}{n}),\\
        \overline{Q}_{t-1}=\overline{Q}_{t-1}-1+B(p\lfloor \frac{P_{2, 
n-\omega_
{1}}}{p+3}\rfloor,\frac{1}{n}).\\
        \end{array}
\right.
\end{equation}
\end{definition}    
  
Let $c = \Pr[(\exists t\in [n-\omega_{0},n-\omega_{1}]): P_{1,t}=0]$. 
Note 
that the amount $p+3$
is subtracted from $X_{t}, Y_{t}, Z_{t}$ exactly once in every
$p$ consecutive steps, so 
whenever the condition (*) is satisfied it holds that $X_{t}\leq
P_{1,t}$ for every $t\in [n-\omega_{0},n-\omega_{1}]$. By coupling
$\Delta_{12,t}^{P}(= B(P_{2,t}, 1/t))$ with $B(P_{2,n-\omega_{1}},1/t)$ 
we
deduce that we can couple $X_{t}$ and $Y_{t}$ so that $Y_{t}\leq
X_{t}$. We can also couple $Y_{t}$ and $Z_{t}$ such that $Z_{t}\leq 
Y_{t}$. 
Finally, notice that we can couple $Z_{n-\omega_{0}-jp}$ and
$\overline{Q}_{n-\omega_{0}-j(p+3)}$ such that
$\overline{Q}_{n-\omega_{0}-j(p+3)}\leq  Z_{n-\omega_{0}-jp}$. 
So an upper bound on $\alpha$ is $\Pr[(\exists t\in [0,n-\omega_{0}]):
\overline{Q}_{t}=0]$. With high probability the Bernoulli distribution
in the definition of the chain $\overline{Q}_{t}$ has the average
strictly  greater than
one, (because the flow from $P_{2,t}$ is approximately Poisson), and 
$\overline{Q}_{n-\omega_{0}}=\Omega(\omega_{0})$, 
therefore, by an elementary property of the queuing chain, the 
probability that $\overline{Q}_t$ hits state 0 is exponentially
small. This yields the desired conclusion, that $\alpha =o(1)$.   
 
One word about the way to prove the fact that $\Phi_{n-\omega_{1}}$ is
unsatisfiable (if defined): one can prove that w.h.p. both
$P_{1,n-\omega_{1}}$ and $N_{1,n-\omega_{1}}$ are
$\Omega(\omega_{1})$. By the uniformity lemma ~\ref{k:2:recurrence} 
we are left with the following instance of the occupancy problem:
there are 
$P_{1,n-\omega_{1}}$ white balls, $N_{1,n-\omega_{1}}$ black balls and
$n-\omega_{1}$ bins. The desired fact now follows from the second part
of Lemma~\ref{occupancy}.

\section{Proof of Theorem~\ref{k:3etc}}

Theorem~\ref{k:3etc} is proved along lines very similar to the proof of
Theorem~\ref{k:2}. The basis is the following generalization of
Lemma~\ref{k:2:recurrence}: 
\vspace{5mm}

\begin{lemma}\label{k:3etc:recurrence}
Suppose that $\Phi$ survives up to stage $t$. Then, conditional on the
values 
$(P_{1,t}, N_{1,t},\ldots, P_{k,t}, N_{k,t})$, the clauses in  
$\Phi_{1,t}^{P},
\Phi_{1,t}^{N},\ldots,  \Phi_{k,t}^{P}, \Phi_{k,t}^{N}$ are chosen 
uniformly
at random and are independent. Also, conditional on the
fact that $\Phi$ survives stage $t$ as well, the following recurrences
hold:
\begin{equation}\label{k:3etc:markovchain}
\left \{\begin{array}{l}
         P_{1,t-1}=P_{1,t}-1-\Delta_{01,t}^{P}+\Delta_{12,t}^{P}, \\        
         N_{1,t-1}=N_{1,t}+\Delta_{12,t}^{N},                    \\
         
P_{i,t-1}=P_{i,t}-\Delta_{0i,t}^{P}-\Delta_{(i-1)i,t}^{P}+\Delta_{i(i+1
),t}^{P}\mbox{, for }i=\overline{2,k},  \\
         
N_{i,t-1}=N_{i,t}-\Delta_{(i-1)i,t}^{N}+\Delta_{i(i+1),t}^{N}\mbox{,\ \
 \  for }i=\overline{2,k},  \\
        \end{array}
\right.
\end{equation}
where (in distribution)
\begin{equation}\label{k:3etc:distribution}
\left \{\begin{array}{l}
\Delta_{01,t}^{P} =B(P_{1,t}-1,1/t),\\
\Delta_{(i-1)i,t}^{P}=B(P_{i,t},(i-1)/t),\\
\Delta_{0i,t}^{P}=B(P_{i,t}-\Delta_{(i-1)i,t}^{P},1/t),\\
\Delta_{(i-1)i,t}^{N}=B(N_{i,t},i/t),\\
\Delta_{k(k+1),t}^{P}=\Delta_{k(k+1),t}^{N}=0.\\
\end{array}
\right.
\end{equation}
\end{lemma}
\vspace{5mm}

\beginproof

The uniformity condition and the justification of the recurrences are 
absolutely similar to the ones from Lemma~\ref{k:2}. 
The additional technical complication is that now there is a ``positive 
flow 
into $P_{2,t}, N_{2,t}$.'' 
\endproof
\qed
\vspace{5mm}

\begin{lemma}
With high probability it holds that 

\[
P_{i,t}=(1+o(1))\cdot \frac{c}{n}\cdot \lambda_{k}\cdot i\cdot 
{{t}\choose
{i}}\cdot S^{n+1-t}_{k-i},
\]
 and 
\[
N_{i,t}=(1+o(1))\cdot \frac{c}{n}\cdot \lambda_{k}\cdot {
{t}\choose
{i}}\cdot S^{n+1-t}_{k-i},
\]
for every $i\geq 2$, and uniformly on $t=n-o(n)$. 
\end{lemma}
\vspace{5mm}

\beginproof

Let us first heuristically derive a formula for $x_{i,t}$, $y_{i,t}$, 
the expected values of $P_{i,t}$, $N_{i,t}$,
obtained by replacing the binomial distributions in the equations by
their expected values. 

We have: 
\begin{equation}\label{k:3etc:markovchain:avg}
\left \{\begin{array}{l}
                 x_{i,t-1}=x_{i,t}- 
\frac{x_{i,t}}{t}-\frac{(i-1)x_{i,t}}{t}+\frac{ix_{i+1,t}}{t}\mbox{, for }i=\overline{2,k},  \\
         
y_{i,t-1}=y_{i,t}-\frac{iy_{i,t}}{t}+\frac{(i+1)y_{(i+1),t}}{t}\mbox{,\ \
 \  for }i=\overline{2,k},  \\
        \end{array}
\right.
\end{equation}
Rearranging terms the recurrences become 
\begin{equation}\label{k:3etc:markovchain:avg:simple}
\left \{\begin{array}{l}
                 
x_{i,t-1}=x_{i,t}(1-\frac{i}{t})+x_{i+1,t}\frac{i}{t}\mbox{, for }i=\overline{2,k},  \\
         
y_{i,t-1}=y_{i,t}(1-\frac{i}{t})+y_{(i+1),t}\frac{(i+1)}{t}\mbox{,\ \
 \  for }i=\overline{2,k}. \\
        \end{array}
\right.
\end{equation}
Also, 
\begin{equation}\label{k:3etc:markovchain:begin}
\left \{\begin{array}{l}
         x_{i,n}= \frac{i{{n}\choose{i}}}{H_{k}}\cdot
         c\lambda_{k}\cdot \frac{H_{k}}{n}=
         \frac{c}{n}\lambda_{k}\cdot i{{n}\choose{i}},\\
         y_{i,n}= \frac{{{n}\choose{i}}}{H_{k}}\cdot c\lambda_{k}\cdot
         \frac{H_{k}}{n}= \frac{c}{n}\lambda_{k}\cdot
         {{n}\choose{i}}.\\
\end{array}
\right.
\end{equation}
A simple induction shows that these expected
values are $x_{i,t}= \frac{c}{n}\cdot \lambda_{k}\cdot i\cdot 
{{t}\choose
{i}}\cdot S^{n+1-t}_{k-i}$, and $y_{i,t}= \frac{c}{n}\cdot 
\lambda_{k}\cdot {
{t}\choose
{i}}\cdot S^{n+1-t}_{k-i}$.

The concentration property can be proved inductively, starting from
$i=k$ towards $3$, by noting that the expected values of the
binomial terms in the recurrence are
$\omega(n)$, hence, by the Chernoff bound,
 the probabilities that they
significantly deviate from their expected values is exponentially
small. 

Almost the same argument holds for 
$P_{2,t}$ and for $N_{2,t})$. 
The only amounts to be handled differently are ``the
clause flows out of $P_{2,t}, N_{2,t}$,'' but they are approximately
Poisson distributed, hence ``small'' with high probability by 
Proposition~\ref{chernoff:poisson}. Therefore $P_{2,t}=(1+o(1)) \frac{c}{n}\cdot
\lambda_{k}\cdot 2\cdot {{t}\choose {2}}\cdot S^{n+1-t}_{k-2}$. 
\endproof
\qed 

The previous lemma implies that $\Delta_{2,t}^{P}\sim Po(c\cdot
\lambda_{k}\cdot S^{n+1-t}_{k-2})$ (for $t=n-o(n)$); thus in this range 
$P_{1,t-1}\sim P_{1,t}-1+Po(c\cdot
\lambda_{k}\cdot S^{n+1-t}_{k-2})$. The proof follows exactly the same
pattern as in the case $c<3/2$ for $k=2$: the conclusion for the stages
$[n,n-\omega_{0}]$ is that the probability that $P_{1.t}$ is zero 
somewhere in this range differs by $o(1)$ from the corresponding
probability for the queuing chain in (\ref{eq:3etc}). The fact that the
stages after $[n,n-\omega_{0}]$ have a contribution of $o(1)$ to the
final accepting probability can be seen by the fact that there is
possible to couple the Markov $M_{1}$, describing the evolution of
\PUR\ on a random $k$-SAT formula, and $M_{2}$ that runs on the  2-CNF
component of the formula, such that for every $t$ we have 
$P_{1,t}^{M_{2}}\leq
P_{1,t}^{M_{1}}$. Perhaps the most intuitive way to see this coupling
is to ``paint'' the initial clauses of the formula having size at most
two in red, and the other clauses in blue. At every step $t$
$P_{1,t}^{M_{2}}$ will count only red clauses having unit size at step
$t$, while $P_{1,t}^{M_{1}}$ will count clauses of both colors. 

Given the stochastic domination, the desired result follows from the
corresponding proof in the case $k=2$.  
\qed

\section{Proof of Proposition~\ref{annealed}}

The idea of the proof is to consider \PUR\ on a random at-most-$k$-Horn 
formula $\Phi$ with 
$\hat{c}\cdot \frac{H_{k}}{n}$ clauses and prove that there exists a
function $\phi(k)$ with $\lim_{k\goesto \infty}\phi(k)=0$ such
that 
\[
\lim_{n\goesto \infty} 
\Pr[\PUR\ \mbox{ accepts in at least } k\mbox{ steps }]\leq \phi(k). 
\]
Indeed, from the previous proof it follows that $\lim_{n\goesto
\infty}\Pr[\PUR\ \mbox{ accepts in }\geq k\mbox{ steps }]$ satisfies 
the recurrence:
\[\label{rec}
x_{t+1}= x_{1,t}-1+Po(\hat{c}\cdot S^{t+1+k}_{k-2}),
\]
where
\[
x_{0}= P_{1,k}\geq 1.
\]
We define $\phi(k)$ to be the probability that the sequence in the
recurrence (\ref{rec}) hits zero. Trivially $\lim_{k\goesto
\infty} S^{k+1}_{k-2}=\infty$, so the expected values of the Poisson
distributions in (\ref{rec}) can be made larger than any given constant
$\lambda$. Using the fact that the sum of two Poisson distributions
with parameters $a$ and $b$ has a Poisson distribution with parameter
$a+b$ it follows that, for large enough $k$, one can couple $x_{t}$ 
with the  
queuing chain
\[\label{rec2}
y_{t+1}= y_{1,t}-1+Po(\lambda),
\]

\[
y_{0}= 1,
\]
such that $y_{t}\leq x_{t}$. It follows that, for large $k$,
$\phi(k)\leq \Pr[\mbox{ the chain $y_{t}$ hits state
zero}]$. Since $\lambda$ was arbitrary, it follows that $\lim_{k\goesto
\infty}\phi(k)=0$. 

Now consider a random {\em uniform} Horn formula $\Phi$ with 
$\hat{c}\cdot \frac{H_{n}}{n}$ clauses, and let $\overline \Phi$ be
its subformula consisting of clauses of size at most $k$. It is easily 
seen
that the behavior of \PUR\ on the first $k-1$ steps depends only on
the clauses of $\overline \Phi$, so 
\[
\Pr[\PUR\ \mbox{ accepts }\Phi\mbox{ in less than }k\mbox{ 
steps}]=\Pr[\PUR\ \mbox{ accepts }\overline
 \Phi\mbox{ in less than }k\mbox{ steps}].
\]
On the other hand we have

\[0\leq \Pr[\PUR\ \mbox{ accepts }\Phi\mbox{ in at least }k\mbox{ 
steps}]\leq \Pr[\PUR\ \mbox{ accepts }\overline
 \Phi\mbox{ in at least }k\mbox{ steps}].
\]
The fact that ``$\overline
 \Phi$ is close to a random formula in $\Omega(n,k,c\cdot
 \frac{H_{k}}{n})$'' (see the discussion in 
Observation~\ref{obs:coupling})
implies that
the right-hand side
term can be made less than any fixed constant $\epsilon$ (for $n,k$
big enough). It follows that 

\[
|\Pr[\PUR\ \mbox{ accepts }\Phi]-\Pr[\PUR\ \mbox{ accepts }\overline
 \Phi]|\leq 2\cdot \epsilon,
\]
for large enough values of $n,k$. This immediately implies the desired 
result.
\endproof
\qed

\section{Proof of Theorem~\ref{k:2:runtime}}
Theorem~\ref{k:2:runtime} is based on the
proof of the Theorem~\ref{k:2} 
and an elementary property of the queuing chain $Q_{t}$
(the expected time to hit state zero, conditional on actually hitting
it has the desired form). 

The crucial point is to prove that the probabilities that any of the
conditions we have employed in our analysis fails have a negligible
effect on the running time.

This is easy to see for stages smaller than $n-\omega_{0}$: since the
probabilities that the various steps of the analysis  
 are either exponentially small or can be made $o(1/n)$ (by choosing a
large enough $k$ in Lemma~\ref{trick:delta},  
the probability that $P_{1,t}$ hits state zero after
stage $n-\omega_{0}$ is $o(1/n)$, therefore its influence on
the average running time of \PUR\ is $o(1)$. 
The corresponding observation  is not true for stages before
$n-\omega_{0}$, but these stages can be handled directly, using the
statement from 
Lemma~\ref{distance}. 
 
\endproof
\qed

\section{Random Horn satisfiability as a mean-field 
approximation}\label{section:4}

What we have shown so far is to prove that (under a suitably rescaled
picture) the rescaled probability graphs for random at-most-$k$ Horn
satisfiability converge to the graph for random Horn satisfiability. 
To be able to argue that our results display critical behavior, we
have to be able to show that this latter probability $p_{\infty}$, 
is indeed the one predicted by some  mean-field
approximation.

In the sequel 
we will show that this is indeed the case. However the mean-field
approximation is {\em not} the one from \cite{kirkpatrick:selman:scaling}
, and incorporates a correction specific to the
properties of random Horn satisfiability. 

Let us
first see that it is not accurate if no correction is taken into
account. Indeed, were it true we would have 

\[ 
\lim_{n\goesto \infty} Pr[\Phi\in \HSAT]= 1 - \lim_{n\goesto
\infty}\prod_{A\in \{0,1\}^{n}} \left(1-\Pr[A \models \Phi]\right). 
\]
Since, for an assignment $A$ of Hamming weight $i$ there are exactly
$2^{i}-1+(n-i)\cdot 2^{i}$ Horn clauses that $A$ falsifies, we have

\[
\Pr[A \models \Phi]= \left(1 - \frac{2^{i}-1+(n-i)\cdot 
2^{i}}{(n+2)\cdot
2^{n}-1}\right)^{c\cdot 2^{n}},
\]
so the mean-field prediction reads

\[\lim_{n\goesto \infty} \Pr[\Phi\in \HSAT]=1- \lim_{n\goesto 
\infty}\prod_{j=0}^{n}\left(1-\left(1 - \frac{2^{j}-1+(n-j)\cdot 
2^{j}}{(n+2)\cdot 2^{n}-1}\right)^{c\cdot 2^{n}}\right)^{{{n}\choose {j}}}. 
\]

All terms in the product are less than 1. Since the term corresponding
to $j=1$ is $\left(1-\left(1 - \frac{1+2\cdot (n-1)}{(n+2)\cdot
2^{n}-1}\right)^{c\cdot 2^{n}}\right)^{n}$ has limit 0, the mean-field 
prediction
would imply that $\lim_{n\goesto \infty} \Pr[\Phi\in \HSAT]=1$.
On the other hand let us observe that, if we do not consider 
the power ${{n}\choose
{j}}$ in the infinite product we obtain the right
result: it is a simple but tedious task to prove that

\[\lim_{n\goesto \infty} \prod_{j=0}^{n}
\left(1-\left(1 - \frac{2^{j}-1+(n-j)\cdot 2^{j}}{(n+2)\cdot 
2^{n}-1}\right)^{c\cdot
2^{n}}\right)=  \prod_{j=0}^{\infty}\left(1-e^{-c \cdot 2^{j}}\right). 
\]

Intuitively  this means that  ``there exist a correction of the 
 mean-field approximation that only considers a single assignment of 
each
 weight, and is accurate.'' The following simple result gives a
 precise statement to the above intuition: 

\begin{lemma}
Suppose $\Phi$ is given as a union of 
formulas $\Phi_{1}, \ldots, \Phi_{n}$, where $\Phi_{i}$ contains all
clauses of length {\em exactly} $i$. Then there is a set 
$T=\{T_{0}, \ldots, T_{n-1}\}$ of assignments, with {\em $T_{i}$ of 
Hamming
weight exactly $i$ and depending
only on $\Phi_{1}\cup \ldots \cup \Phi_{i+1}$}, 
such that $\Phi$ is satisfiable if and only if it is
satisfied by some assignment in $T$. 
\end{lemma}

\beginproof

Let $\overline{y_{1}\ldots y_{k}}$ denote the assignment that makes 
$y_{1}=\ldots = y_{k}=1$, and all the other variables equal to zero. 

The set $T$ has two parts: the first is simply the set of 
assignments implicitly examined by the algorithm \PUR\ in testing
satisfiability. That is, if $x_{1}, \ldots, x_{k}$ are the variables
assigned by \PUR\  in this order, the first part includes the
assignments $00000$, $\overline{x_{1}}, \ldots,\overline{x_{1},\ldots,
x_{k}}$. The second part contains a random assignment for each 
remaining weight. 
\endproof
\qed
The result has a ``mean-field'' interpretation: as before, define 
$f(x_{1}, \ldots, x_{n})= 1-
\prod_{i=1}^{n} x_{i}$, and the function 
$g_{k}[\Phi]$ to be  the indicator function for the event ``$T_{k}
\not \models \Phi$, given that event $\overline{A}_{n}\AND \ldots \AND
\overline{A}_{n-k+1}$ happens,'' i.e. 

\[g_{k}[\Phi]= \frac{1}{\Pr[\overline{A}_{n}\AND \ldots \AND
\overline{A}_{n-k+1}]}\cdot \left \{\begin{array}{ll}
                 1, & \mbox{ if } T_{k} \not \models \Phi \AND 
\overline{A}_{n}\AND \ldots \AND
\overline{A}_{n-k+1}
                 \\
 
                 0,  & \mbox{ otherwise.}\\ 
        \end{array}
\right.
\]
We have 

\[
E[g_{k}[\Phi]]= \Pr[\overline{A_{n-k}}|\overline{A}_{n}\AND \ldots \AND
\overline{A}_{n-k+1}].
\]
Indeed, $g_{k}[\Phi]\neq 0$ exactly when $R_{n}\OR \ldots \OR
R_{n-k+1}$ or $T_{k}\not \models \Phi \AND S_{n}\AND
\ldots S_{n-k+1}$. The second event is equivalent to
$\overline{A_{n-k}}\AND S_{n}\AND \ldots S_{n-k+1}$, hence we have 
$g_{k}[\Phi]\neq 0$ exactly when $\overline{A_{n-k}}\AND
\overline{A}_{n}\AND \ldots \AND\overline{A}_{n-k+1}$ holds. 

Thus we have, by the discussion in the previous chapter, 
\[ 
f(E[g_{1}[\Phi]], \ldots,E[g_{n}[\Phi]])= 
1-\prod_{k=0}^{n}\Pr[\overline{A_{n-k}}|\overline{A}_{n}\AND \ldots \AND
\overline{A}_{n-k+1}]= \Pr[\Phi \in \HSAT]. 
\]

The above correction seems
to be specific to the random model for Horn satisfiability, which
allows clauses of varying lengths.

To sum up: {\em the mean-field approximation is true, modulo a
correction that takes into account some particular features of the
random model for Horn satisfiability}. 

\section{Discussion}

We have characterized the asymptotical
satisfiability probability of a random $k$-Horn formula, and 
showed that it exhibits very similar behavior to the one uncovered
experimentally in \cite{kirkpatrick:selman:scaling}. 

We have also displayed an ``easy-hard-easy'' pattern similar to the
ones observed experimentally in the AI literature. In our case the 
pattern is fully explained by elementary properties of the queuing 
chain. 

As for an explanation of the ``critical behavior'', 
consider an intermediate stage $i$ of \PUR\ and 
let $C_{j}$ be the set of clauses of $\Phi_{i,j}^{P}$.  
It is clear that whether \PUR\ accepts is
dependent only on the number of clauses in $C_{1}$. The
restriction on the clause length acts like a ``dampening''
perturbation (in that it eliminates the ``clause flow into $C_{k}$''). 
The proof of Theorem~\ref{k:infinite} states that when
 $k(n)\goesto \infty$, with high probability \PUR\ accepts (if
$\Phi$ is satisfiable) ``before the perturbation reaches $C_{1}$'', 
therefore the satisfiability probability is the one from the uniform
case. On the other hand, for any constant $k$,  with probability
greater than 0 \PUR\ does not 
halt during the first $k$ iterations (for the exact value see 
\cite{istrate:cs.DS/9912001}), and the dampening has a
significant influence. Thus {\em the explanation for  
the occurrence (and specific form of) critical behavior is a threshold 
property for the number of iterations of \PUR\ on random satisfiable 
Horn
formulas ``in the critical region''}.

A related, and somewhat controversial, open issue is whether random
Horn satisfiability properly displays critical behavior. Problems with 
a sharp threshold display ``critical'' (i.e singular) behavior at least 
in one parameter, the satisfaction probability, which conceivably 
allows the definition of critical exponents. This is not so for random 
$k$-Horn satisfiability, that has a coarse 
threshold, and no criticality for $k>2$, hence the question seems not 
to be 
meaningful. Note, however, that the order parameter involved in the 
recent study of the phase transition in 2-SAT \cite{scaling:window:2sat} 
is {\bf not} satisfaction probability, but the (expected size) of the 
so-called {\em backbone} (or its more tractable version {\em spine}) of a 
random formula. The ``window'' that we use to peek at the threshold 
behavior of random 
Horn satisfiability does not seem to be ``naturally required'' by any 
physical 
considerations, and it is possible in principle that the random 
Horn formulas display critical behavior if we take the spine as the 
order parameter.

\section{Acknowledgments}

This paper is part of the author's Ph.D. thesis at the University of
Rochester. 
Support for this work has come from the
NSF CAREER Award CCR-9701911 and the
NSF Grant 9725021.

{\small
\bibliography{/home/gistrate/bib/bibtheory}
\clearpage
 }

\end{document}